\documentclass[11pt,onecolumn,english,showpacs,preprintnumbers,amsmath,amssymb,floatfix,nofootinbib]{revtex4-1}

\usepackage{epsfig}
\usepackage{graphics}
\usepackage{latexsym}
\usepackage{amsmath}
\usepackage{amssymb}
\usepackage{rotating}
\usepackage{subfigure}
\usepackage{bm}
\usepackage{color}
\usepackage{ulem}
\usepackage{booktabs}
\usepackage[colorlinks = true,
linkcolor = magenta,
urlcolor  = blue,
citecolor = red,
anchorcolor = blue]{hyperref}

\usepackage{extsizes}
\usepackage{geometry}
\usepackage{pxfonts}
\usepackage{graphicx}
\usepackage{booktabs}
\usepackage[table]{xcolor}
\definecolor{lightgray}{gray}{0.85}
\usepackage{color}

\begin{document}

%
%
\title{ Impact of unidentified light charged hadron data on the determination of pion fragmentation functions } 
%
%

\author{Maryam Soleymaninia$^{1,3}$}
\email{Maryam\_Soleymaninia@ipm.ir}

\author{Muhammad Goharipour$^{3}$}
\email{Muhammad.Goharipour@ipm.ir}

\author{Hamzeh Khanpour$^{2,3}$}
\email{Hamzeh.Khanpour@mail.ipm.ir}

\affiliation {
$^{1}$Institute of Advanced Technologies, Shahid Rajaee Teacher Training University, Lavizan, Tehran, 16788, Iran          \\
$^{2}$Department of Physics, University of Science and Technology of Mazandaran, P.O.Box 48518-78195, Behshahr, Iran       \\
$^{3}$School of Particles and Accelerators, Institute for Research in Fundamental Sciences (IPM), P.O.Box 19395-5531, Tehran, Iran
 }


%
\begin{abstract}

In this paper a new comprehensive analysis of parton-to-pion fragmentation functions (FFs) is performed for the first time by including all experimental data sets on single inclusive pion as well as unidentified light charged hadron production in electron-positron ($e^+e^-$) annihilation.
We determine the pion FFs along with their uncertainties using the standard ``Hessian" technique at next-to-leading order (NLO) and next-to-next-to leading order (NNLO) in perturabative QCD. It is shown that the determination of pion FFs using simultaneously the data sets from pion and unidentified light charged hadron productions leads to the reduction of all pion FFs uncertainties especially for the case of strange quark and gluon FFs by significant factors. In this study, we have quantified the constraints that these data sets could impose on the extracted pion FFs.
Our results also illustrate the significant improvement in the precision of FFs fits achievable by inclusion of higher order corrections. The improvements on both FFs uncertainties as well as fit quality have been clearly discussed.

\end{abstract}

\maketitle
\tableofcontents{}

%
\section{Introduction}\label{sec:introduction}
%

Essential ingredients of theoretical predictions for the present or future hadron colliders such as the large hadron collider (LHC) and large hadron-electron collider (LHeC), are the detailed understanding of the quark and gluon structure of the nucleon~\cite{Khalek:2018mdn,Gao:2017yyd,Rojo:2015acz,Forte:2013wc}. These are quantified by the parton distribution functions (PDFs)~\cite{Ball:2018odr,Ball:2017nwa,Alekhin:2017kpj} as well as the fragmentation functions (FFs)~\cite{Bertone:2018ecm,Bertone:2017tyb,Soleymaninia:2017xhc,Soleymaninia:2018uiv,Mohamaditabar:2018ffo,Hirai:2016loo,MoosaviNejad:2016qdx,Nejad:2015fdh,MoosaviNejad:2017rvi,Soleymaninia:2013cxa,Boroun:2016zql,Boroun:2015aya,Zarei:2015jvh}. In recent years, precise determination of PDFs as well as FFs including their experimental uncertainties had become an active topic for many LHC processes, including top-quark and Higgs boson sector, searches for new heavy beyond the Standard Model (BSM) particles, searches for new physics (NP) as well as in the measurement of fundamental SM parameters such as the strong coupling constant. For more details, we refer the readers to the literature~\cite{deFlorian:2016spz,Alioli:2017jdo,Ball:2018iqk} and a recent study on the PDFs at the High-Luminosity LHC (HL-LHC)~\cite{Khalek:2018mdn}.

In a hard-scattering collision, PDFs determine how the proton's momentum is shared among its constituents. Likewise, the FFs describe the probability density for the fragmentation of the final-state parton with a certain momentum into the hadron with a fraction of the parton's momentum. 
PDFs and the FFs depend on the factorization scale. This dependence is described by the DGLAP evolution equations~\cite{Gribov:1972ri,Lipatov:1974qm,Altarelli:1977zs,Dokshitzer:1977sg}, which allow the calculation
of the PDFs and FFs, if they are known at a given initial scale, i.e. $\mu^2 = \mu_0^2$. 
It is well known that the PDFs and the FFs can not be calculable in perturbation theory, and hence, these distributions need to be extracted from experimental
information through a QCD fit. In addition, these non-perturbative functions are also universal. The universality of PDFs and FFs commonly refers that, since the hadronization processes are not sensitive to the particular choices of hard scattering process in short range, these non-perturbative functions can be extracted from certain kind of scattering experimental observables. Then the extracted distributions can be used for the theory predictions of scattering observable in high energy collisions.

The new and precise data sets are vital for the precise determination of FFs. These data sets have been and currently been collected from different high energy processes at variety of lepton and hadron colliders. These processes include the hadron production data in single-inclusive electron-positron ($e^+e^-$) annihilation (SIA), semi-inclusive deep inelastic scattering (SIDIS), and proton-proton and proton-antiproton collisions measured by TEVATRON, RHIC and LHC. For a list of all available data sets, we refer the readers to the recent analysis by NNPDF collaboration and references therein~\cite{Bertone:2018ecm,Bertone:2017tyb}.
Several analyses have been done so far to extract FFs using the observables mentioned above. Among them are the recent determination of charged hadron FFs from collider data by NNPDF collaboration, {\tt NNFF1.1h}~\cite{Bertone:2018ecm}. This collaboration also have determined the pions, kaons, and proton FFs using the SIA data sets at NNLO in perturbative QCD based on the NNPDF methodology, {\tt NNFF1.0}~\cite{Bertone:2017tyb}.
The recent analyses by {\tt HKKS16}~\cite{Hirai:2016loo} and {\tt JAM16}~\cite{Sato:2016wqj} also have been performed using the SIA data only. Other analyses in literature can be found for example in Refs.~\cite{Anderle:2017cgl,Albino:2005mv,Albino:2008fy,Aidala:2010bn,Kretzer:2000yf,Kniehl:2000fe,Bourhis:2000gs,deFlorian:2007ekg}

Recently, we also have performed the First determination of $D^{*\pm}$-meson FFs and their uncertainties at NNLO, {\tt SKM18}~\cite{Soleymaninia:2017xhc}. In Ref.~\cite{Soleymaninia:2018uiv} we presented our QCD analysis of charged hadron FFs and their uncertainties at NLO and NNLO ({\tt SGK18}) which is the first determination of light charged hadron FFs at NNLO accuracy. Finally in Ref.~\cite{Mohamaditabar:2018ffo} the contributions from {\it residual} light charged hadrons in the inclusive charged hadrons have been extracted using the $e^+e^-$ annihilation data sets. Since the QCD framework for FFs at NNLO are not accessible for SIDIS, and hadron-hadron collisions, both of our analyses are restricted to the single-inclusive charged hadron production in electron-positron annihilation. The uncertainties in our recent analyses on FFs as well as the corresponding observables are estimated using the ``Hessian'' technique.

In this work, an extraction of pion FFs from QCD analysis of electron-positron annihilation experimental data in zero-mass variable flavor number scheme (ZM-VFNS) has been presented. The main aim of this paper is to examine, for the first time, the impact of unidentified light charged hadron experimental data on the determination of pion FFs and their uncertainties at NLO and NNLO accuracy.
In this respect, we have attempted a determination of pion FFs considering two different scenarios.
First, we present a determination of pion FFs through a QCD analysis of pion data sets. In this first study of FFs, which is performed within ZM-VFNS at both NLO and NNLO approximations and referred to as ``{\tt pion fit}'', we simplify the analysis by considering the pion data sets only. Secondly, we determine pion FFs through a QCD analysis by including both pion and unidentified light charged hadron data sets. We show that the fitting simultaneously the pion FFs using both data sets leads to a well-constrained determination of pion FFs including significant effect on the extracted uncertainties. Our second fit entitled as ``{\tt pion+hadron fit}''.

The outline of this paper is as follows.
In section~\ref{sec:data selection}, we present in details all available SIA data sets for pion production as well as the SIA data sets for the unidentified light charged hadrons.
In Section~\ref{sec:QCD analysis}, we discuss the theoretical formalism of single-hadron inclusive production in electron-positron ($e^+ e^-$) annihilation. 
This section also includes the detailed discussions of our fitting process and parameterization for the pion FFs. Section~\ref{sec:results} is then dedicated to our results. The obtained results are clearly discussed for variety of aspect in this section, and comparison with other analyses in literature also presented.
This section also includes our theory predictions based on the extracted pion FFs including a comparison with all data analyzed. Finally, Section~\ref{sec:conclusion} includes a summary and our conclusions.

%
\section{ Experimental data selection } \label{sec:data selection}
%

In this section, we present the experimental data sets that are included in our ``{\tt pion fit}'' and ``{\tt pion+hadron fit}'' analyses. As we mentioned in the Introduction, our QCD fits are performed by inducing the electron-positron annihilation data in two scenarios:
In the first analysis, we use the available SIA data for pion from Refs.~\cite{Buskulic:1994ft,Abreu:1998vq,Akers:1994ez,Brandelik:1980iy, Althoff:1982dh,Braunschweig:1988hv,Leitgab:2013qh,Lees:2013rqd,Aihara:1988su,Abe:2003iy} to extract the pion FFs. In the second analysis, the SIA data sets for the unidentified charged hadrons~\cite{Buskulic:1995aw,Ackerstaff:1998hz,Akers:1995wt,Abreu:1998vq,Abreu:1997ir,Aihara:1988su,Abe:2003iy,Braunschweig:1990yd} along with the pion data sets are included in our fits to calculate the FFs of pion.
All the data sets for pion and unidentified hadrons are listed in Tables.~\ref{tab:datasetsNLO} and \ref{tab:datasetsNNLO} for inclusive and flavor-tagged SIA data which are reported by different experiments.
Note that, the measured observables for these data sets, specially for pion, are different and a complete explanation about SIA pion data and the relations between the scaling variables are available in related analysis done by NNPDF collaboration in {\tt NNFF1.0}~\cite{Bertone:2017tyb}. In addition, we have used the unidentified light charged hadron experimental data in our recent study of ({\tt SGK18})~\cite{Soleymaninia:2018uiv}. The details of corrections to these data sets and the kinematic cuts applied are presented in Ref.~\cite{Soleymaninia:2018uiv}.  

According to the data sets presented in second column of Tables.~\ref{tab:datasetsNLO} and \ref{tab:datasetsNNLO}, the observables are different and provide limited sensitivity to the separation between light and heavy quark FFs due to the flavor-tagged data. Since the gluon receives its leading order (LO) accuracy at $\cal{O}(\alpha_s)$, the total SIA cross sections are poor to constrain this density. However, the longitudinal cross sections can impose a comparable sensitivity to the gluon FF because the longitudinal coefficient functions start at $\cal{O}(\alpha_s)$. Hence, the longitudinal observables that are available for the unidentified hadrons could constrain the gluon FF well enough. It should be noted that the NNLO QCD corrections for longitudinal structure functions are not available in the literature, and hence, such corrections can not be considered in our analyses.

In this paper, we plan to study the effects arising from the unidentified light charged hadron experimental data on the calculation of pion FFs by including both pion and unidentified hadron data sets, and then, compare the extracted pion FFs with the results calculated from the QCD analysis using pion data sets alone. Since the most contribution of FFs into the unidentified light charged hadron cross sections mainly comes from the identified pion FFs, it motivates us to investigate the effect of unidentified light charged hadron data sets on the reduction of pion FFs uncertainties. In Tables.~\ref{tab:datasetsNLO} and \ref{tab:datasetsNNLO}, our results are reported at NLO and NNLO accuracies of perturbative QCD. In both tables, the forth column presents our fit results for the value of $\chi^2$ per number of data points ($\chi^2/N_{pts.}$) considering pion data sets in the fit, while in the fifth column the same quantity are reported considering the pion and light hadron experimental data sets in the analysis. One of the most important findings from these tables are the significant reduction of $\chi^2/{dof}$ by going from NLO to the NNLO corrections. We will return to this issue in the next section.

In order to avoid the sensitivity of behaviors of FF parametrization in the low and high regions of $z$, we apply cuts on the momentum fraction $z$. We exactly follow the cuts applied in our recent study on light charged hadron FFs, {\tt SGK18}~\cite{Soleymaninia:2018uiv}. These selections are also imposed for the pion experimental data. For data sets at $\sqrt{s}=M_Z$, we include the data points with the scaling variable of $z\geq0.02$ and for $\sqrt{s}<M_{Z}$, the data points with $z \geq 0.075$ are included in our QCD fits. The data points with $z>0.9$ are excluded in all of our QCD analyses. Considering the kinematical cut applied, the number of the data points are listed separately in the denominator of the forth and fifth columns in Tables.~\ref{tab:datasetsNLO} and \ref{tab:datasetsNNLO} for the NLO and NNLO accuracy, respectively.

\begin{table}
\begin{center}
\tiny
\rowcolors{2}{lightgray}{}
\resizebox*{\textwidth}{!}{
\begin{tabular}{lp{2.2cm}p{1.6cm}p{2.2cm}p{2.2cm}}
				\hline 
				Dataset     & observable    & $\sqrt{s}$~[GeV]    &  $\chi ^2/{pts.}$ ``pion" & $\chi ^2/{pts.}$ ``pion+hadron" \\ 
				\hline 
				BELLE~\cite{Leitgab:2013qh}  & inclusive & 10.52  & 38.37/70& 42.28/70  \\ 
				BABAR~\cite{Lees:2013rqd} & inclusive & 10.54 & 78.07/40 & 81.64/40 \\ 
				TASSO12~\cite{Brandelik:1980iy} & inclusive & 12 & 4.24/4 &4.05/4 \\ 
				TASSO14~\cite{Althoff:1982dh}& inclusive & 14 & 11.76/9& 12.04/9  \\       					
				TASSO22~\cite{Althoff:1982dh} & inclusive & 22 & 25.39/8 & 26.55/8 \\ 
				TPC~\cite{Aihara:1988su}& inclusive & 29 & 7.05/13  &8.25/13 \\ 
				TASSO34~\cite{Braunschweig:1988hv}& inclusive& 34 & 19.22/9&  23.26/9\\ 
				TASSO44~\cite{Braunschweig:1988hv}& inclusive & 44 & 18.21/6 &19.95/6  \\ 
				ALEPH~\cite{Buskulic:1994ft}& inclusive & 91.2 & 37.77/23 &43.07/23 \\ 
				DELPHI~\cite{Abreu:1998vq} & inclusive & 91.2 & 27.52/21 &22.86/21 \\ 
				& $uds$ tag & 91.2 & 21.47/21  &22.70/21\\ 
				& $b$ tag &91.2 & 21.12/21  & 11.11/21\\ 
				OPAL~\cite{Akers:1994ez}& inclusive & 91.2 & 32.01/24& 37.41 /24 \\ 
				SLD~\cite{Abe:2003iy} & inclusive & 91.2 & 57.87/34 &76.20/34\\ 
				& $uds$ tag &91.2 & 90.98/34 & 92.04/34\\ 
				& $c$ tag& 91.2 & 38.83/34  & 40.13/34\\ 
				& $b$ tag&91.2 & 19.81/34  &38.28/34\\ 
				TASSO14~\cite{Braunschweig:1990yd} & inclusive & 14 & ----& 8.22/15  \\ 
				TASSO22~\cite{Braunschweig:1990yd} & inclusive & 22 & ---- & 13.07/15 \\ 
				TPC~\cite{Aihara:1988su}& inclusive & 29 & ----- &20.72/21 \\ 
				TASSO35~\cite{Braunschweig:1990yd}& inclusive& 34 & ----&  21.74/15\\ 
				TASSO44~\cite{Braunschweig:1990yd}& inclusive & 44 & ----&18.80/15  \\ 
				ALEPH~\cite{Buskulic:1995aw}& inclusive & 91.2 & ---- &9.23/32 \\ 
				DELPHI~\cite{Abreu:1998vq,Abreu:1997ir} & inclusive & 91.2 & ---- &16.86/22 \\ 
				& $uds$ tag & 91.2 & ----  &10.52/22\\ 
				& $b$ tag &91.2 & ----  & 51.76/22\\ 
				& Longitudinal inclusive & 91.2 & ----  &28.49/20\\ 
				& Longitudinal $b$ tag &91.2 & ----  & 20.12/20\\ 
				OPAL~\cite{Ackerstaff:1998hz,Akers:1995wt} & inclusive & 91.2 & ----& 12.40/20  \\ 
				& $uds$ tag & 91.2 & ----  &7.34/20\\ 
				& $c$ tag &91.2 & ----  & 14.18/20\\ 
				& $b$ tag & 91.2 & ----  &26.85/20\\ 
				&Longitudinal inclusive &91.2 & -----  & 89.18/20\\ 
				SLD~\cite{Abe:2003iy} & inclusive & 91.2 & ----- &15.09/34\\ 
				& $uds$ tag &91.2 & ---- & 15.86/34\\ 
				& $c$ tag& 91.2 & ----  & 29.26/34\\ 
				& $b$ tag&91.2 & -----  &81.21/34\\ 				
				\rowcolor{white}				\hline
				Total $\chi^2$/{dof}  &  &  & 1.42 & 1.44   \\ 
				\hline
\end{tabular} 		}
\end{center}
\caption{ \small The data sets included in the analyses of $\pi^\pm$ FFs at NLO. For each experiment, we indicate the corresponding reference, the measured 
observables, the center-of-mass energy $\sqrt{s}$, the $\chi^2/{pts.}$ values for every data set, as well as the total $\chi^2/{dof}$.  }
\label{tab:datasetsNLO}	
\end{table}

\begin{table}
\begin{center}
\tiny
\rowcolors{2}{lightgray}{}
\resizebox*{\textwidth}{!}{
\begin{tabular}{lp{2.2cm}p{1.6cm}p{2.2cm}p{2.2cm}}
				\hline 
				Dataset     & observable    & $\sqrt{s}$~[GeV]    &  $\chi ^2/{pts.}$ ``pion" & $\chi ^2/{pts.}$ ``pion+hadron" \\ 
				\hline 
				BELLE~\cite{Leitgab:2013qh} & inclusive & 10.52  & 27.39/70& 29.96/70  \\ 
				BABAR~\cite{Lees:2013rqd} & inclusive & 10.54 & 59.84/40 & 57.80/40 \\ 
				TASSO12~\cite{Brandelik:1980iy} & inclusive & 12 & 4.28/4 &4.21/4 \\ 
				TASSO14~\cite{Althoff:1982dh}& inclusive & 14 & 11.50/9& 11.67/9  \\       					
				TASSO22~\cite{Althoff:1982dh} & inclusive & 22 & 23.17/8 & 24.09/8 \\ 
				TPC~\cite{Aihara:1988su}& inclusive & 29 & 10.07/13  &9.26/13 \\ 
				TASSO34~\cite{Braunschweig:1988hv}& inclusive& 34 & 14.44/9&  15.93/9\\ 
				TASSO44~\cite{Braunschweig:1988hv}& inclusive & 44 & 16.93/6 &17.78/6  \\ 
				ALEPH~\cite{Buskulic:1994ft}& inclusive & 91.2 & 27.63/23 &35.50/23 \\ 
				DELPHI~\cite{Abreu:1998vq} & inclusive & 91.2 & 29.79/21 &24.78/21 \\ 
				& $uds$ tag & 91.2 & 22.22/21  &23.57/21\\ 
				& $b$ tag &91.2 & 19.96/21  & 10.57/21\\ 
				OPAL~\cite{Akers:1994ez}& inclusive & 91.2 & 30.53/24& 35.74 /24 \\ 
				SLD~\cite{Abe:2003iy} & inclusive & 91.2 & 37.60/34 &47.80/34\\ 
				& $uds$ tag &91.2 & 68.97/34 & 66.70/34\\ 
				& $c$ tag& 91.2 & 31.73/34  & 35.18/34\\ 
				& $b$ tag&91.2 & 19.36/34  &40.38/34\\ 
				TASSO14~\cite{Braunschweig:1990yd} & inclusive & 14 & ----& 8.78/15  \\ 
				TASSO22~\cite{Braunschweig:1990yd} & inclusive & 22 & ---- & 13.22/15 \\ 
				TPC~\cite{Aihara:1988su}& inclusive & 29 & ----- &15.69/21 \\ 
				TASSO35~\cite{Braunschweig:1990yd}& inclusive& 34 & ----&  23.33/15\\ 
				TASSO44~\cite{Braunschweig:1990yd}& inclusive & 44 & ----&19.41/15  \\ 
				ALEPH~\cite{Buskulic:1995aw}& inclusive & 91.2 & ---- &10.62/32 \\ 
				DELPHI~\cite{Abreu:1998vq,Abreu:1997ir} & inclusive & 91.2 & ---- &18.55/22 \\ 
				& $uds$ tag & 91.2 & ----  &11.66/22\\ 
				& $b$ tag &91.2 & ----  & 50.99/22\\ 
				& Longitudinal inclusive & 91.2 & ----  &9.47/20\\ 
				& Longitudinal $b$ tag &91.2 & ----  & 9.37/20\\ 
				OPAL~\cite{Ackerstaff:1998hz,Akers:1995wt} & inclusive & 91.2 & ----& 14.23/20  \\ 
				& $uds$ tag & 91.2 & ----  &8.53/20\\ 
				& $c$ tag &91.2 & ----  & 14.56/20\\ 
				& $b$ tag & 91.2 & ----  &26.41/20\\ 
				&Longitudinal inclusive &91.2 & -----  & 7.99/20\\ 
				SLD~\cite{Abe:2003iy} & inclusive & 91.2 & ----- &10.31/34\\ 
				& $uds$ tag &91.2 & ---- & 10.97/34 \\ 
				& $c$ tag& 91.2 & ----  & 29.74/34 \\ 
				& $b$ tag&91.2 & -----  &80.62/34 \\ 				
				\rowcolor{white}				\hline
				Total $\chi^2$/{dof}  &  &  & 1.17&1.06   \\ 
				\hline
\end{tabular}		}
\end{center}	
\caption{ \small Same as Table.~\ref{tab:datasetsNLO} but at NNLO accuracy. }
\label{tab:datasetsNNLO}
\end{table}

%
%
\section{ Theoretical methodology for calculations and fitting }\label{sec:QCD analysis}
%

In this section, a brief review of the theoretical framework and our methodology has been presented. 
According to the factorization theorem, the SIA differential cross section normalized to the total cross section $\frac{1}{\sigma_{\rm tot}}
\frac{d \sigma^{\it H^\pm}} {dz}$ at a given center-of-mass energy of $\sqrt{S}=Q$  is written by,
\begin{equation}\label{cross}
\frac{1}
{\sigma_{\rm tot}}
\frac{d 
\sigma^{\it  H^{\pm}}}
{dz} =
\frac{1}
{\sigma_{\rm tot}}
\left[ F_T^{\it H^\pm}
(z, Q) +
F_L^{\it H^\pm}
(z, Q) \right] \,.
\end{equation}
This equation is used for identified charged hadrons such as $\pi^\pm$, $K^\pm$ and $p/\bar{p}$ and, unidentified hadrons $h^\pm$. In Eq.~\eqref{cross}, $H^\pm$ is defined as sum of different charge of hadrons $H = H^+ + H^-$ and $z=\frac{2 E_H}{\sqrt{s}}$ is the scaling variable. The total cross section $\sigma_{\rm tot}$ depends to the perturbative order of QCD corrections and detail explanations can be found, for example, in Ref.~\cite{Soleymaninia:2018uiv}.
According to the Eq.~\eqref{cross}, in the case of multiplicities, the differential cross section for SIA processes can be decomposed into time-like structure functions $F_T$ and $F_L$ which are the transverse (T) and longitudinal (L) perturbative parts, respectively.
The time-like structure functions can be written as convolutions of a perturbative part, coefficient functions $C_i(z,\alpha_s)$, and a nonperturbative part, FFs $D^{H^\pm}(z, Q)$,
\begin{equation}\label{structure}
F^{H^\pm}(z,Q)
=\sum_i C_i(z,\alpha_s)
\otimes
D^{H^\pm}(z,Q).
\end{equation}
The coefficient functions have been calculated in Refs.~\cite{Rijken:1996ns,Mitov:2006wy,Rijken:1996vr} and they are available up to NNLO accuracy for electron positron annihilations. 
It should be mentioned here that, in this analysis, the renormalization scale $\mu_R$ and the factorization scale $\mu_F$ considered to be equal to the center-of-mass energy of collision, $\mu_R = \mu_F = \sqrt{s}$.

Since the universal FFs are nonperturbative functions, in order to determine the FFs, one needs to parametrize the functions of partons $i=q, \, \bar{q}, \, g$ at a given initial scale. The $z$ parameter represents the fraction of the parton momentum which carried by hadron.
Theoretically, the renormalization equations govern the scale dependence of the FFs and they can be evaluate to a given higher energy scale using the DGLAP evolution equations.
In our analysis, we use the publicly {\tt APFEL} package~\cite{Bertone:2013vaa} in order to calculate of the SIA cross sections as well as the evolution of FFs by DGLAP equations up to NNLO accuracy. In addition, the ZM-VFNS is considered  to account the heavy quarks contributions, and hence, the effects of heavy quark mass are not taken into account in our analysis. 

Our main aim in this analysis is to study the effect of adding all the unidentified light charged hadrons experimental data to the pion ones from SIA processes in the procedure of determination of pion FFs.
Hence, we need the theoretical definition of unidentified charge hadron FFs in our calculations. Experimentally, the unidentified light charged hadrons contain all identified light hadrons such as pion, kaon, proton and a small {\it residual} light hadrons. Then unidentified charged hadron cross sections of SIA can be calculated by summing of individual cross sections of the identified light ones ($\pi^\pm$, $K^\pm$ and $p/\bar{p}$) and the {\it residual} contribution.
The SIA coefficient functions for all final states are the same, and hence, the FFs of unidentified light charged hadrons ($D^{h^\pm}$) can be defined as the sum of the pion, kaon and proton FFs ($D^{\pi^\pm}$, $D^{K^\pm}$, $D^{p/\bar{p}}$) including the {\it residual} light hadron FFs $D^{{res}^\pm}$
\begin{eqnarray}\label{unidentified def}
D^{h^\pm} =
D^{\pi^\pm} + D^{K^\pm}
+ D^{p/\bar{p}} +
D^{{\it res}^\pm}.
\end{eqnarray}
Since our aim in this analysis is a new determination of pion FFs $D^{\pi^\pm}$, we use the kaon and proton FFs from {\tt NNFF1.0} parton set~\cite{Bertone:2017tyb} both at NLO and NNLO accuracies. 
Recently, we have calculated the {\it residual} light hadron FFs $D^{{res}^\pm}$ in Ref.~\cite{Mohamaditabar:2018ffo} up to NNLO QCD correction. In Ref.~\cite{Mohamaditabar:2018ffo}, we have shown that the contribution of the {\it residual} light hadrons are small, and hence, one can ignore this small contribution in Eq.~\eqref{unidentified def}. The contribution from this small distribution are not significant for the case of total or light charged cross sections, however, for the case of $c$- and $b$-tagged cross sections they are sizable.  

For the uncertainty from {\tt NNFF1.0}, we follow the analysis by {\tt DSS07} in Ref.~\cite{deFlorian:2007ekg} and estimate an average uncertainty of 5\% in all theoretical calculations of the inclusive charged hadron cross sections stemming from the large uncertainties of kaon and proton FFs from {\tt NNFF1.0} set. In addition, our recent study shows that an additional  uncertainty due to the contributions of {\it residual} charged hadrons FFs~\cite{Mohamaditabar:2018ffo} also need to be taken into account. Overall, we believe that a 8\% of the cross section value seems to be reasonable. These additional uncertainties are included in the $\chi^2$ minimization procedure for determining the pion FFs.
In order to add these uncertainties, we apply such a simplest way to include a ``theory'' error which we add it in quadrature to the statistical and systematic experimental error in the $\chi^2$ expression. This is the standard approach that one can use to add this additional uncertainty to the QCD analysis. The method of the present study are also consistent with those of {\tt DSS07}~\cite{deFlorian:2007ekg} who used the same approach, and hence, our results share a number of similarities with {\tt DSS07} findings. This method was chosen because it is one of the most practical and economic ways to include such uncertainty and in agreement from previous results reported in the literature. However this method may suffers from a number of pitfalls. One need to use a rigorous approach and include the full {\tt NNFF1.0} uncertainties in the kaon and proton FFs in Eq.~\eqref{unidentified def}. In order to ensure the affect of this alternative method on our conclusions, we also examined this approach. Our study shows that one can reaches the same conclusions, finding no increase in the size of uncertainty.  For the physical parameters, we exactly follow the  analysis by NNFF collaboration, {\tt NNFF1.0}. We use the heavy flavor masses for charm and bottom as $m_c=1.51$~GeV and $m_b=4.92$~GeV~\cite{Bertone:2018ecm,Bertone:2017tyb}, respectively. Also the Z-boson mass is chosen to be $M_Z=91.187$~GeV and the QCD coupling constant is fixed to the world average $\alpha_s(M_Z)=0.1185$~\cite{Tanabashi:2018oca}.

Now we are in a position to present our QCD fit methodology, input functional form as well as the assumptions we used in our analysis to determine the pion FFs.
We choose a flexible input parametrization for pion FFs at initial scale $Q_0$ which we also used in our very recent analysis of unidentified light charged hadrons~\cite{Soleymaninia:2018uiv},
\begin{eqnarray}\label{input}
D^{\pi^\pm}_i(z, Q_0)
= \frac{{\cal N}_i
z^{\alpha_i}(1 - z)
^{\beta_i}
[1 + \gamma_i
(1 - z)^{\delta_i}]}
{B[2 + \alpha_i,
\beta_i + 1] +
\gamma_i B[2 +
\alpha_i, \beta_i +
\delta_i + 1]}, \,
\nonumber \\
\end{eqnarray}
where $i = u^+, d^+ , s^+, c^+$, $b^+$ and $g$, $q^+ = q + \bar{q}$. In order to normalize the parameter ${\cal N}_i$ we use the Euler Beta function $B[a,b]$.
Since we include the FF sets of {\tt NNFF1.0} for kaon and proton, we choose the initial scale of energy $Q_0=5$~GeV and therefore the number of active flavors in our analyses need to be fixed at $n_f=5$.
In addition, the charge conjugation and isospin symmetry $D^{\pi^\pm}_{u^+} = D^{\pi^\pm}_{d^+}$ are assumed. More specifically, the $\gamma$ and $\delta$ parameters for $s^+$, $c^+ $ and $g$ could not well constrain by the SIA data and we are forced to fix them as $\gamma _{s^+, c^+, g}=0$ and $\delta _{s^+, c^+, g}=0$. Then the best fit is only achieved with all five parameters of Eq.~\eqref{input} for $u^+$ and $b^+$. We determine 19 free parameters by a standard $\chi^2$ minimization strategy in which the details can be found in Refs.~\cite{Soleymaninia:2018uiv,Khanpour:2016pph}. 

The free parameters are determined from the best fit, and we list them in Table.~\ref{tab:pars}. In the second and third columns of this table, we report our best fit parameters for only pion data analysis at NLO and NNLO accuracy, respectively. The parameters reported by the forth and fifth columns are for the analyses with both pion and unidentified hadron data sets at both perturbative orders.

\section{ Analysis results } \label{sec:results}

After the detailed presenting of the experimental data sets included in the present work and the theoretical and phenomenological framework of the analysis in the previous sections, in the following we present the numerical results obtained for the pion FFs from different analyses and compare them with each other.
As we mentioned before, the main goal of the present work is to investigate, for the first time, the impact of unidentified light charged hadron experimental data on the pion FFs at both NLO and NNLO accuracy.
In this respect, the pion FFs should be determined by performing two different analyses: 1) determination of pion FFs through a QCD analysis of only pion data sets as usual ({\tt pion fit}), and 2) determination of pion FFs through a simultaneous analysis of both pion and unidentified light charged hadron data sets ({\tt pion+hadron fit}).

The important point that should be noted is the presence of the kaon, proton and {\it residual} FFs in the theoretical calculation of the unidentified light charged hadron cross sections which is required for the second analysis. As discussed in Sec.~\ref{sec:QCD analysis}, we use the kaon and proton FFs from the {\tt NNFF1.0} analysis~\cite{Bertone:2017tyb} and ignore the small {\it residual} contribution. Hence, some theoretical uncertainties should be taken into account in the analysis containing the unidentified light charged hadron data. One of the most common methods is adding a point-to-point uncertainty to the experimental data as a systematic error source, 8\% in our analyses.

\begin{table*}
\begin{center}
\tiny
\rowcolors{2}{lightgray}{}
\resizebox*{\textwidth}{!}{
\begin{tabular}{lp{2.2cm}p{2.2cm}p{2.2cm}p{2.2cm}}
				\hline
				Parameter   & ``pion" NLO  &  ``pion" NNLO & ``pion+hadron" NLO &``pion+hadron" NNLO  \\ 
				\hline
				${\cal N}_{u^+}$& $1.123$& $1.062$& $1.133$&  $1.071$\\ 
				$\alpha_{u^+}$& $-0.617$& $-0.713$& $-0.558$ &$-0.671$ \\ 
				$\beta_{u^+}$ & $1.737$& $1.854$& $1.757$ &$1.862$ \\ 
				$\gamma_{u^+}$ & $8.324$& $6.550$& $9.705$ & $7.742$ \\ 
				$\delta_{u^+}$ & $5.175$& $5.843$& $5.314$& $6.163$  \\ 
				${\cal N}_{s^+}$& $0.239$& $0.456$& $0.124$ & $0.397$ \\ 
				$\alpha_{s^+}$& $1.634$& $0.598$& $3.376$  & $0.986$\\ 
				$\beta_{s^+}$ & $10.714$& $8.468$& $12.658$ & $8.873$ \\ 
				${\cal N}_{c^+}$& $0.739$& $0.777$& $0.724$ & $0.773$ \\ 
				$\alpha_{c^+}$& $-0.903$& $-0.901$& $-0.929$ & $-0.907$ \\ 
				$\beta_{c^+}$ & $4.662$& $5.055$& $4.520$ & $4.917$ \\ 
				${\cal N}_{b^+}$& $0.694$& $0.735$& $0.673$ & $0.735$ \\ 
				$\alpha_{b^+}$& $-0.395$& $-0.446$& $-0.346$ & $-0.449$ \\ 
				$\beta_{b^+}$ & $5.346$& $5.057$& $4.728$  & $4.500$\\ 
				$\gamma_{b^+}$ & $6.014$& $7.356$& $9.098$ & $8.735$ \\ 
				$\delta_{b^+}$ & $9.102$& $8.567$& $10.573$  & $9.086$\\ 
				${\cal N}_{g}$& $0.616$& $0.571$& $0.705$ & $0.611$ \\ 
				$\alpha_{g}$& $0.406$& $0.137$& $-0.230$ & $-0.068$ \\ 
				$\beta_{g}$ & $14.210$& $16.174$& $8.658$ & $13.688$ \\ 	
				\hline \hline			\end{tabular} 		}
\end{center}
\caption{ The best fit parameters for the fragmentation of partons into the $\pi^\pm$ for both {\tt pion fit} and {\tt pion+hadron fit} analyses at NLO and NNLO accuracy. The starting scale is taken to be $Q_{0}=5$~GeV for all parton species. }
\label{tab:pars}	
\end{table*}

\subsection{Comparison of $\chi^2$ values }

The list of experimental data sets including their references as well as the results of our analyses introduced above have been summarized in Tables.~\ref{tab:datasetsNLO} and~\ref{tab:datasetsNNLO} at NLO and NNLO, respectively. In each table, the second column indicates the kind of observable measured by each experiment and the third column specifies its related value of center-of-mass energy. Note also that the columns labeled by ``{\tt pion}" and ``{\tt pion+hadron}" are containing the results of the first and second analyses, respectively.
The values of $\chi^2$ per number of data points ($\chi^2/N_\textrm{pts.}$) have been presented in these columns for each data set. Moreover, the value of total $\chi^2$ divide by the number of degrees of freedom ($\chi^2/\textrm{dof}$) for each analysis is presented in the last raw of the table.
The total number of data points included in the ``{\tt pion fit}" analysis is 405, while it is 879 for the ``{\tt pion+hadron fit}" analysis.
According to the results obtained, the following conclusions can be drawn. For the case of NLO analyses, although the values of $\chi^2/N_\textrm{pts.}$ have increased almost for each pion data set after the inclusion of the unidentified light charged hadron data, but the values of $\chi^2/\textrm{dof}$ for the ``{\tt pion fit}" and ``{\tt pion+hadron fit}" analyses are almost equal. Such behavior is seen for some of the data sets in the case of NNLO analyses, but with the difference that the value of $\chi^2/\textrm{dof}$ has decreased by including the unidentified light charged hadron data in the analysis. Another point should be noted here is the significant reduction in the value of $\chi^2/\textrm{dof}$ when we move from NLO to NNLO. The optimum values of fit parameters have been presented in Table.~\ref{tab:pars}, where the first and second columns are related to the pion data analyses at NLO and NNLO, respectively, while the third and fourth columns contain the results of the simultaneous analyses of the pion and hadron data at NLO and NNLO accuracy.

\subsection{Comparison of the relative uncertainties }

\begin{figure*}[htb]
	\vspace{0.50cm}
	\resizebox{0.48\textwidth}{!}{\includegraphics{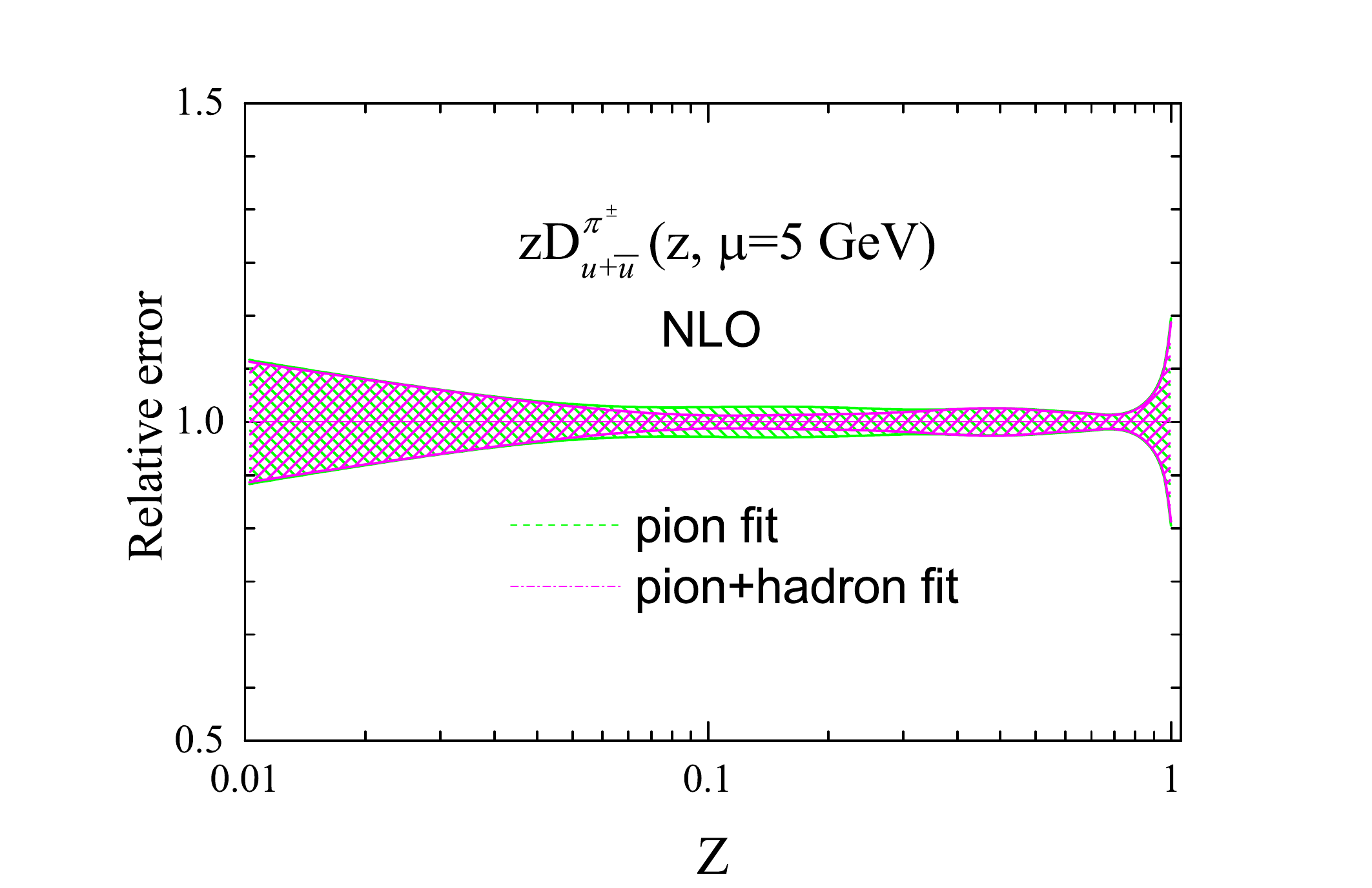}}  
	\resizebox{0.48\textwidth}{!}{\includegraphics{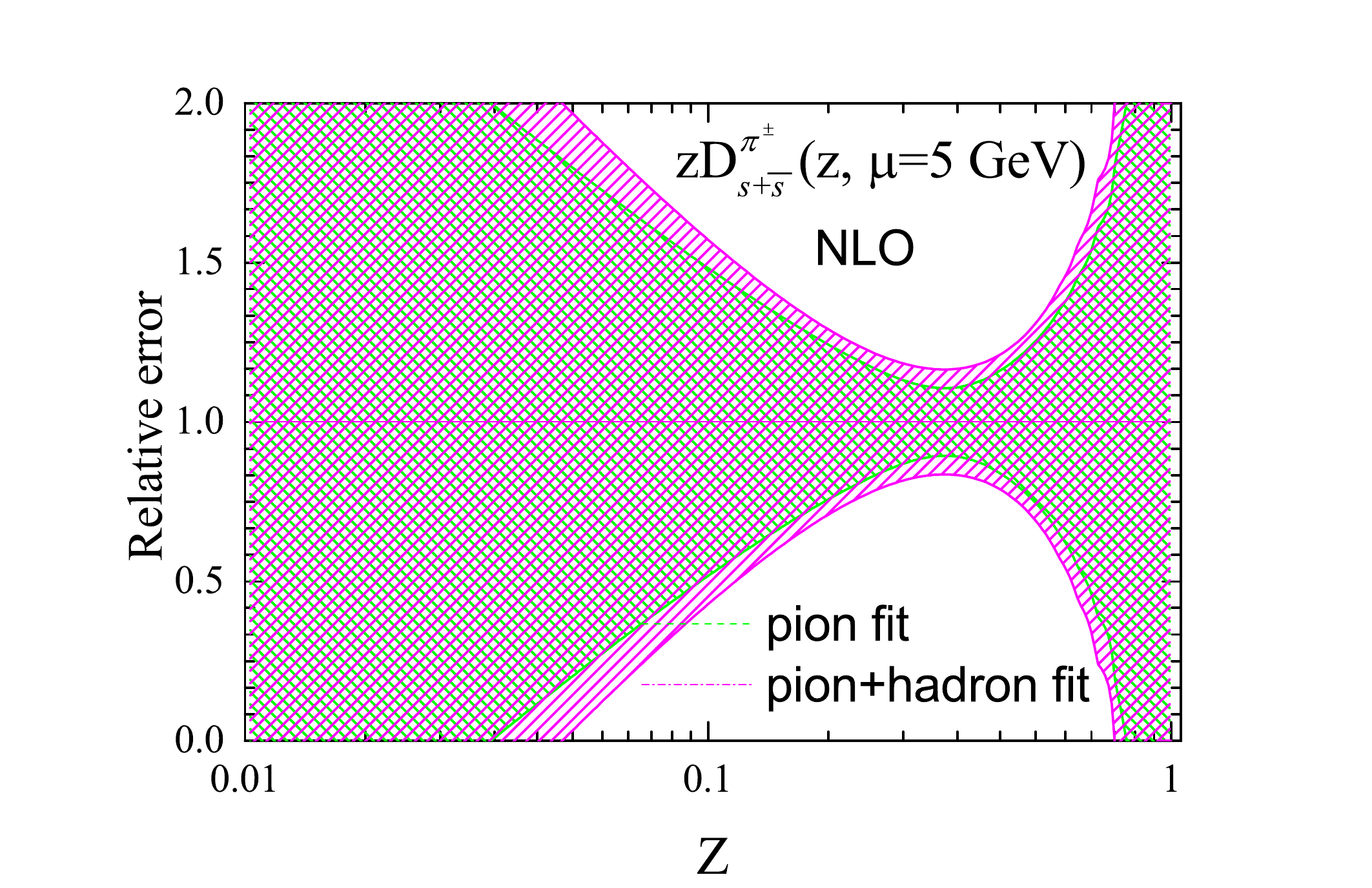}} 
	\resizebox{0.48\textwidth}{!}{\includegraphics{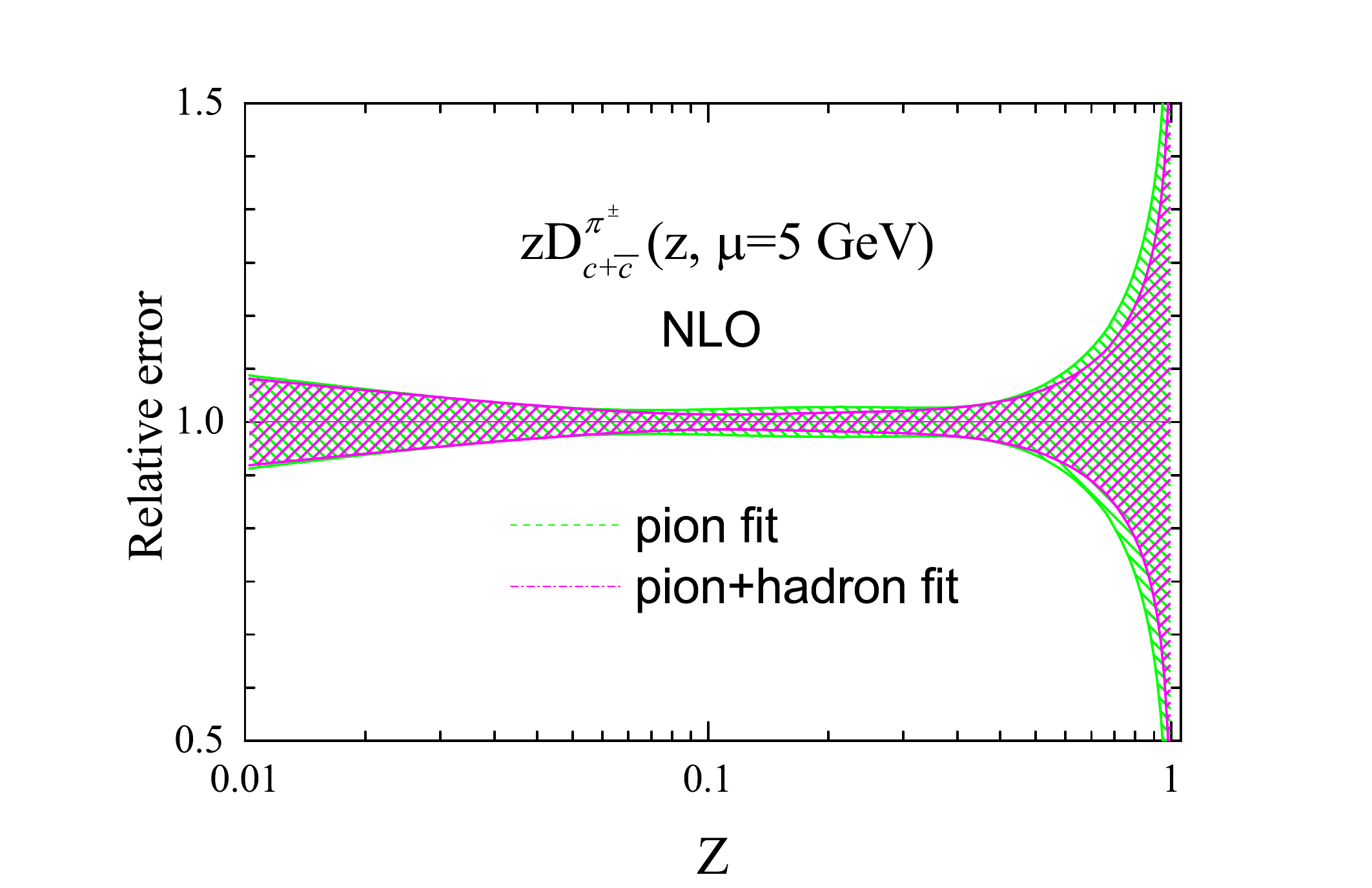}}  
	\resizebox{0.48\textwidth}{!}{\includegraphics{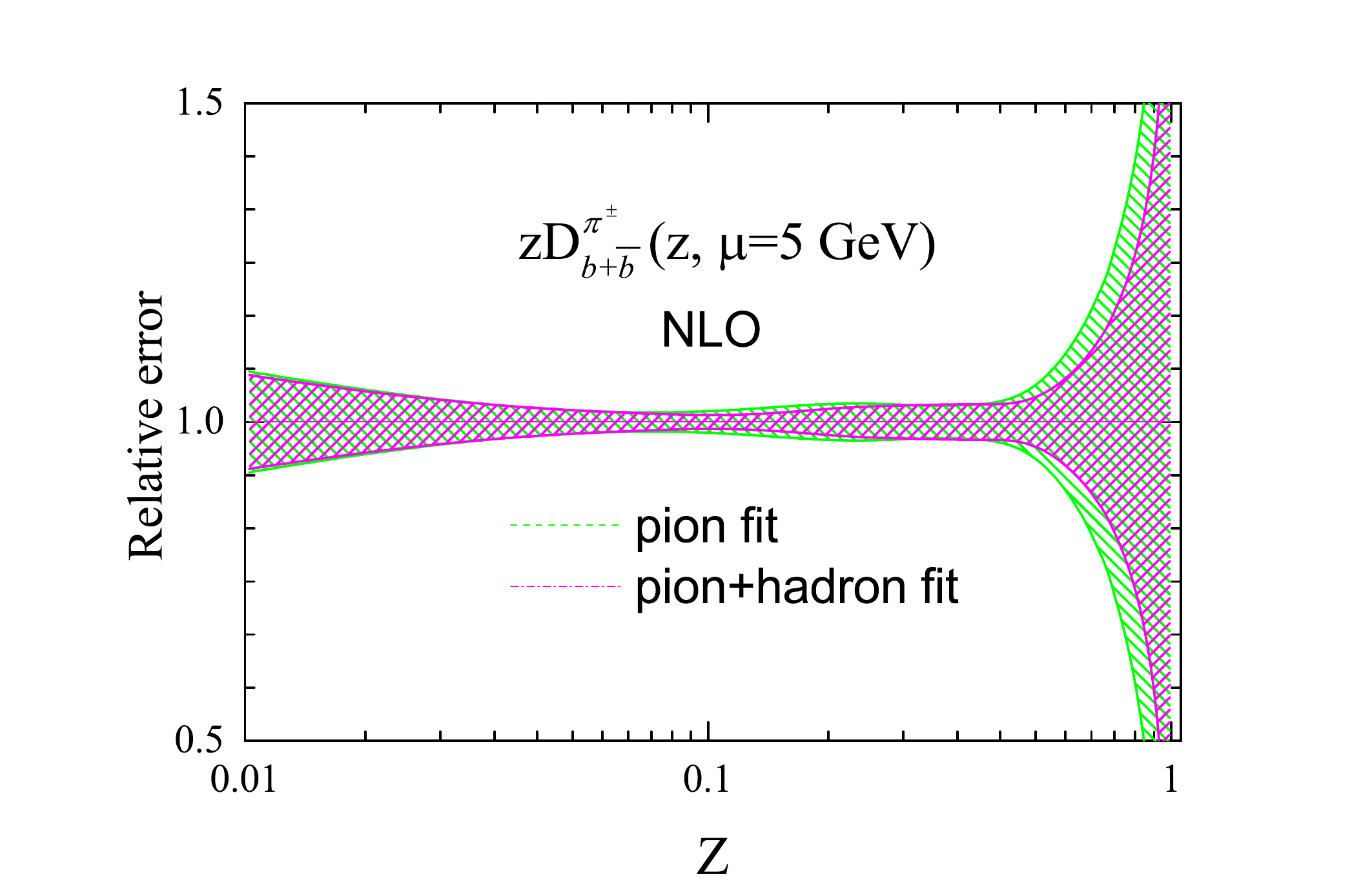}} 
	\resizebox{0.48\textwidth}{!}{\includegraphics{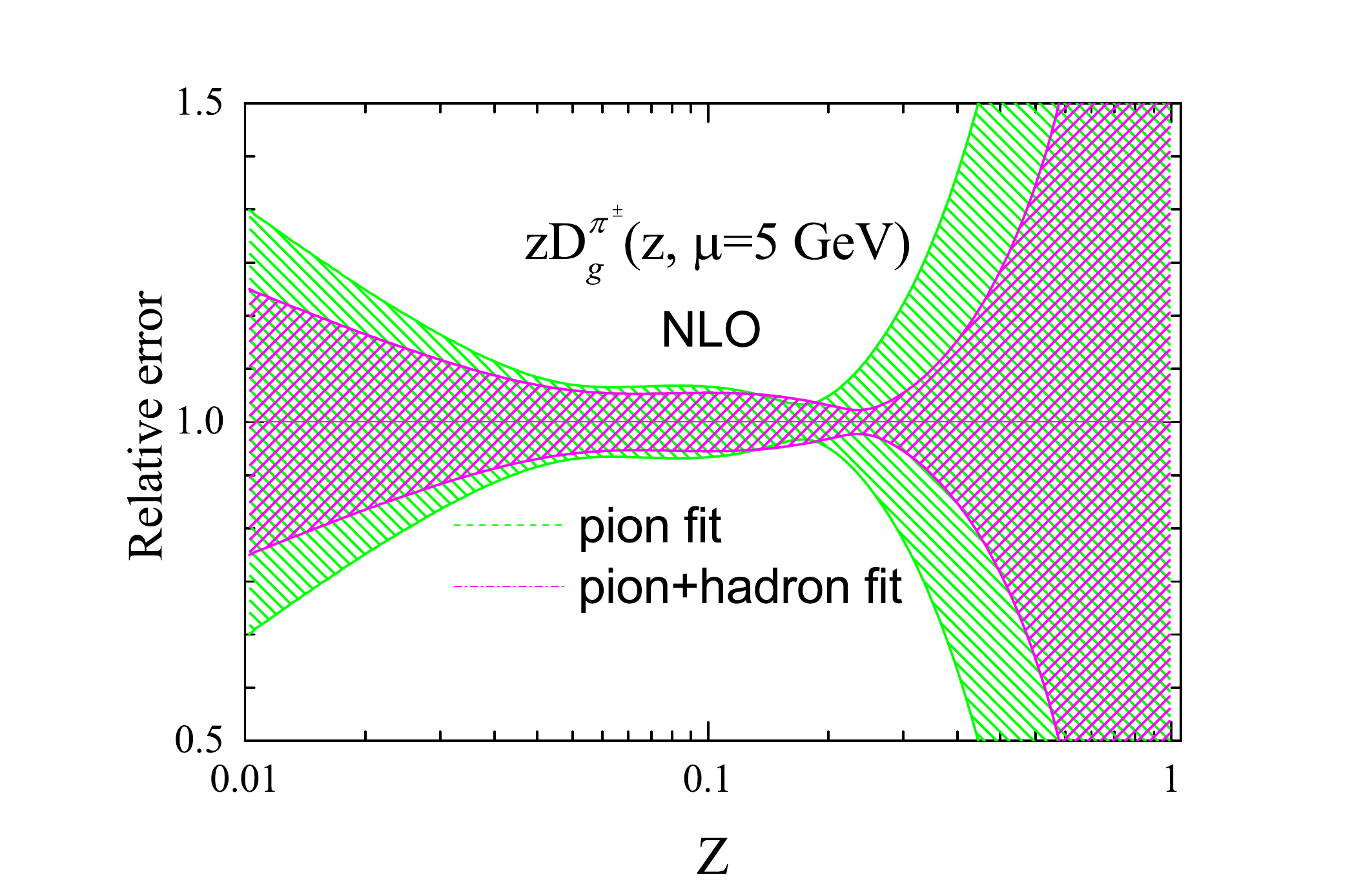}} 
	\begin{center}
		\caption{{\small Comparison between the relative uncertainties of pion FFs at $Q_0=5$~GeV obtained from the ``{\tt pion fit}" and ``{\tt pion+hadron fit}" analyses at NLO. } \label{fig:Ratioq5NLO}}
	\end{center}
\end{figure*}

In order to investigate the impact arising from the inclusion of unidentified light charged hadron experimental data on pion FFs both in behavior and uncertainty, the results obtained from ``{\tt pion fit}" and  ``{\tt pion+hadron fit}" can be compared in various ways. One of the best approaches to check the validity and excellency of the new results obtained, specifically in view of the uncertainties, is comparing the relative uncertainties of the extracted distributions which are obtained, for each analysis separately, by dividing the upper and lower bands to the central values.
Fig.~\ref{fig:Ratioq5NLO} shows a comparison between the relative uncertainties of pion FFs obtained from the ``{\tt pion fit}" and ``{\tt pion+hadron fit}" analyses at NLO accuracy. We have presented the results for all flavors parameterized in the analysis at the initial scale of $Q_{0}=5$~GeV. As can bee seen, except for the case of $s+\bar s$ FF, the relative uncertainties of pion FFs obtained from the simultaneous analysis of the pion and hadron data are smaller than those obtained by fitting the pion data alone, especially for the case of gluon FF. In fact, the amount of the uncertainty of $s+\bar s$ FF from ``{\tt pion+hadron fit}" analysis is also less than ``{\tt pion fit}" analysis (as will be shown later), but since its central value is smaller by a factor of two, it has overall a relative uncertainty which is somewhat larger.

\begin{figure*}[htb]
	\vspace{0.50cm}
	\resizebox{0.48\textwidth}{!}{\includegraphics{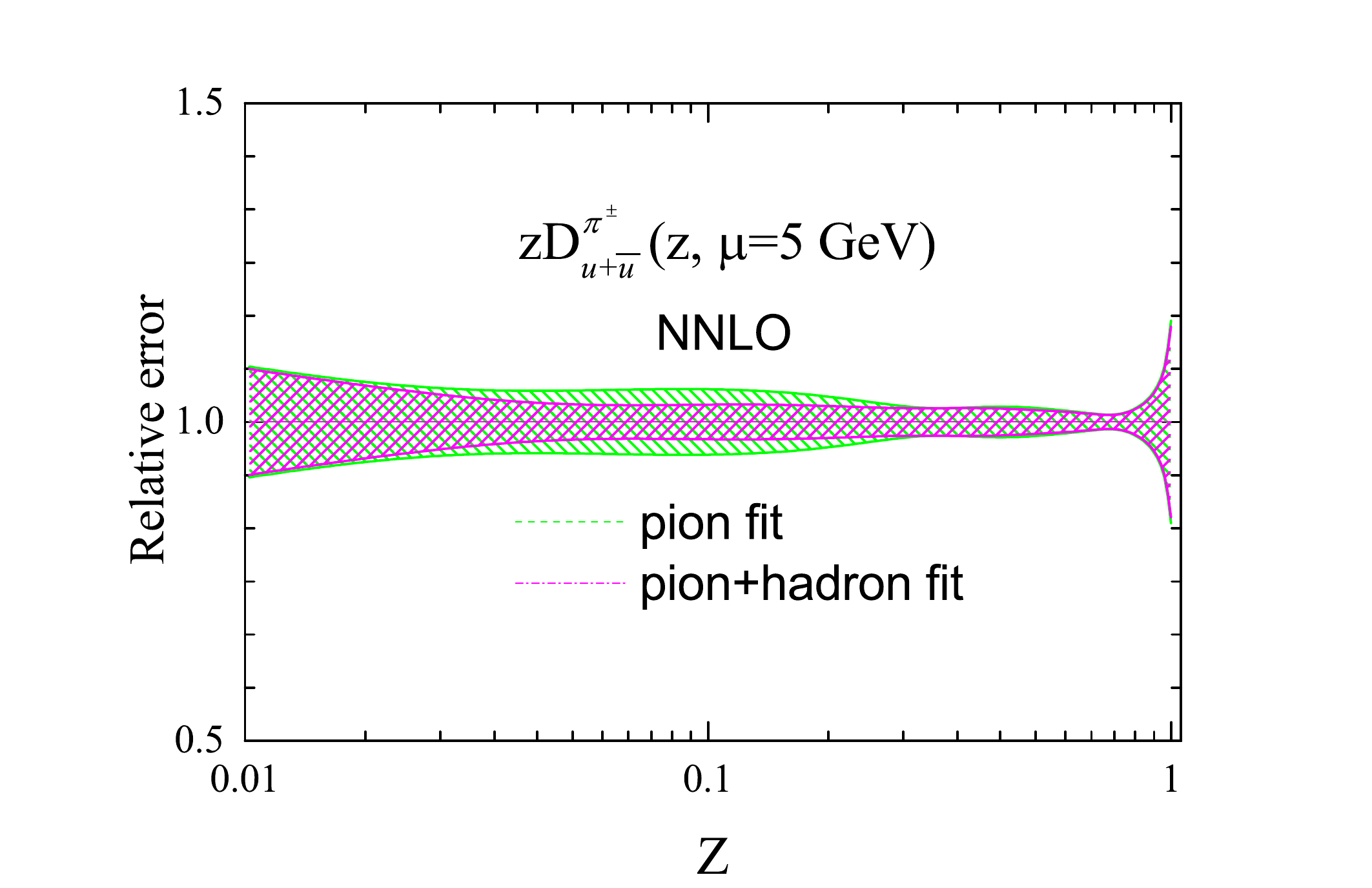}}  
	\resizebox{0.48\textwidth}{!}{\includegraphics{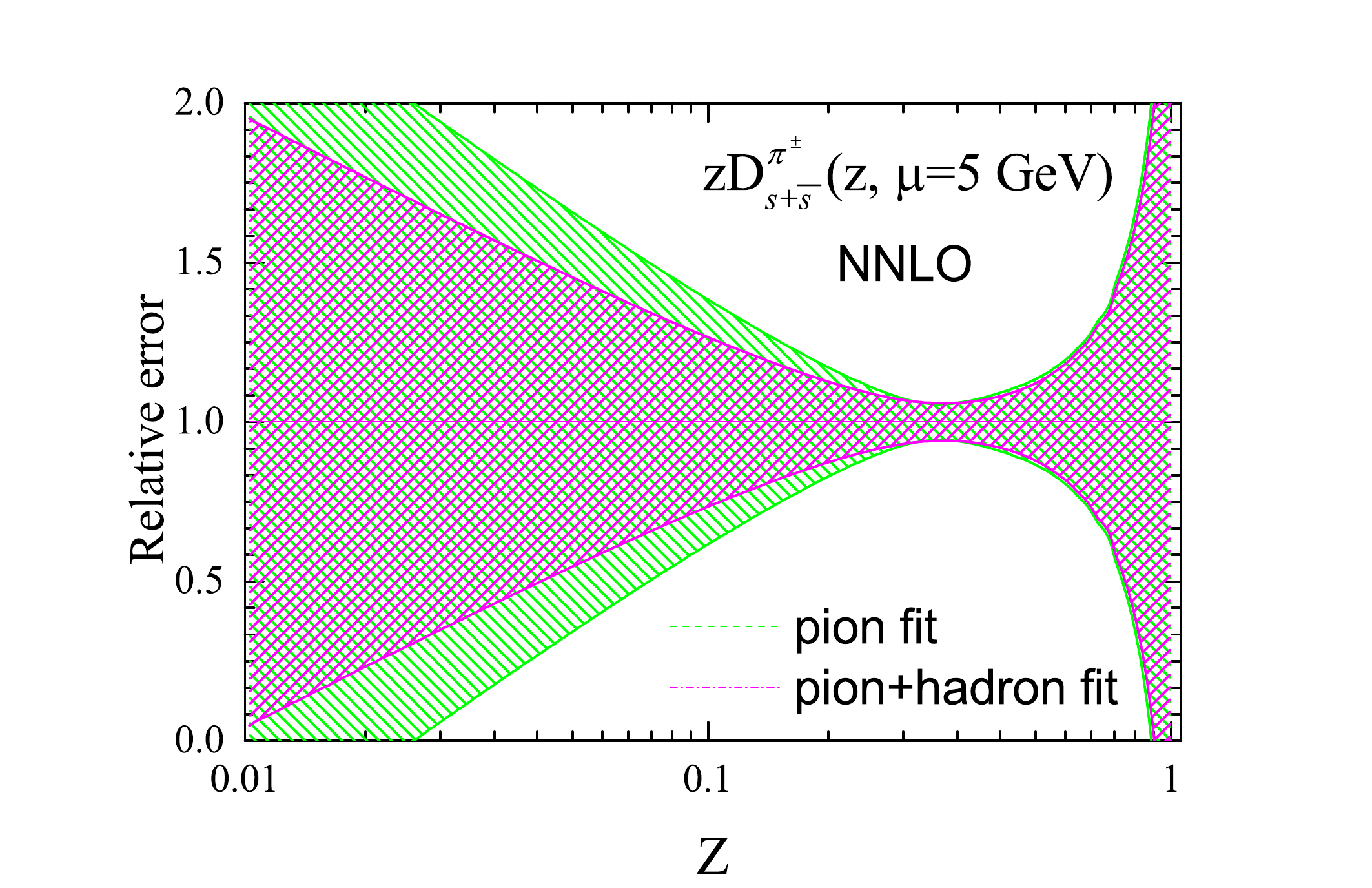}} 
	\resizebox{0.48\textwidth}{!}{\includegraphics{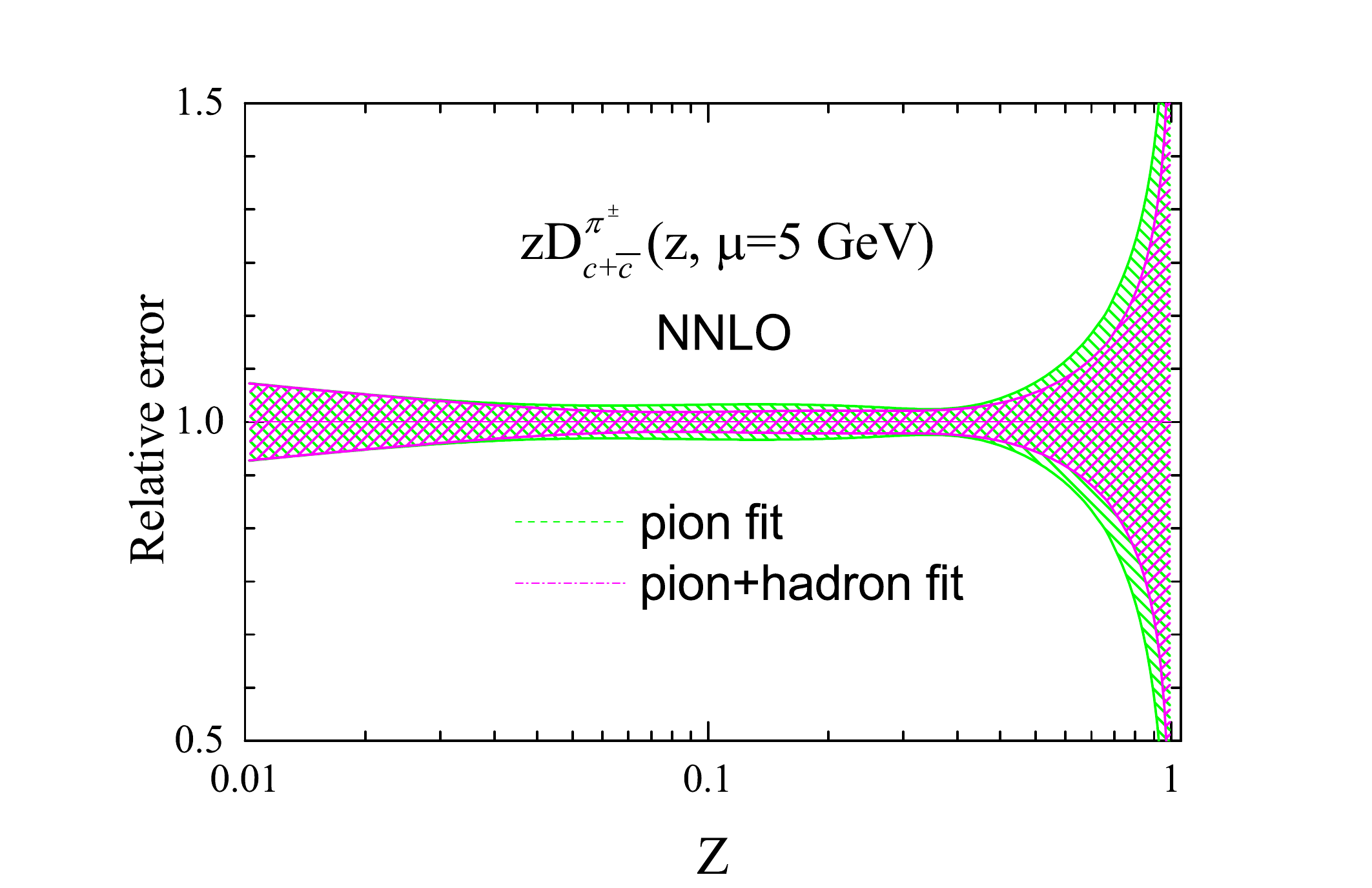}}  
	\resizebox{0.48\textwidth}{!}{\includegraphics{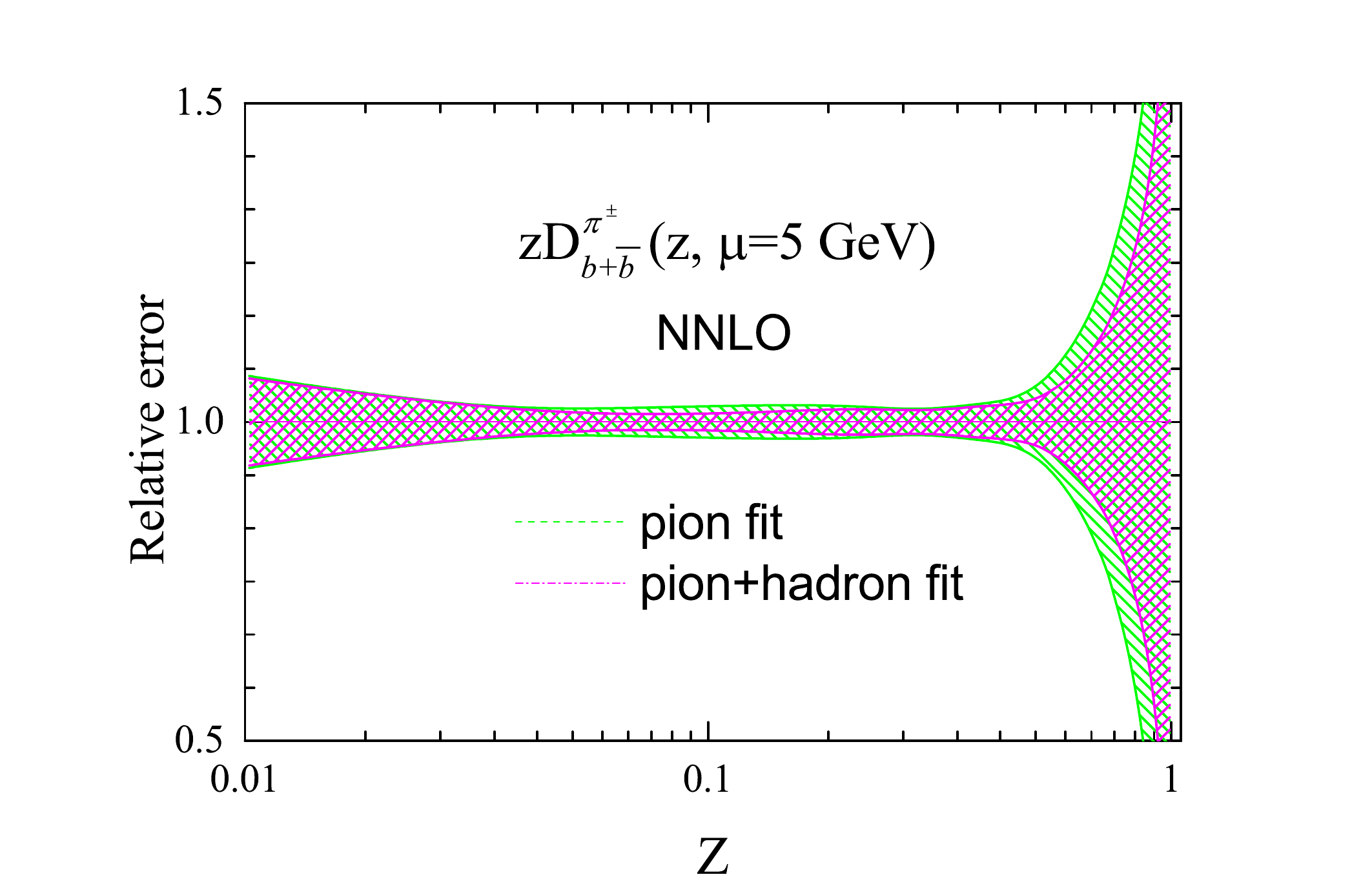}} 
	\resizebox{0.48\textwidth}{!}{\includegraphics{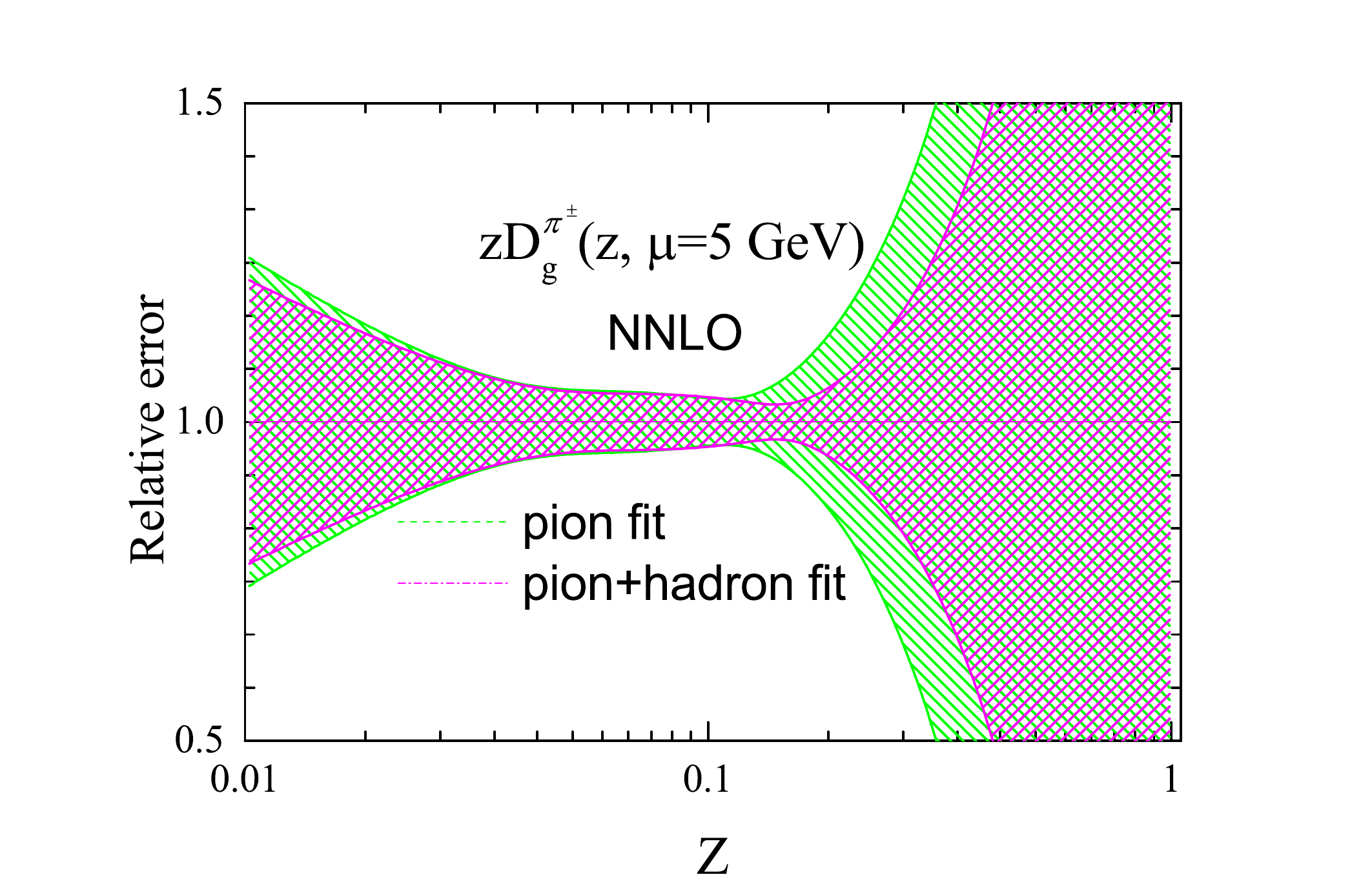}} 
	\begin{center}
		\caption{{\small Same as Fig.~\ref{fig:Ratioq5NLO} but at NNLO. } \label{fig:Ratioq5NNLO}}
	\end{center}
\end{figure*}

\begin{figure*}[htb]
	\vspace{0.50cm}
	\resizebox{0.48\textwidth}{!}{\includegraphics{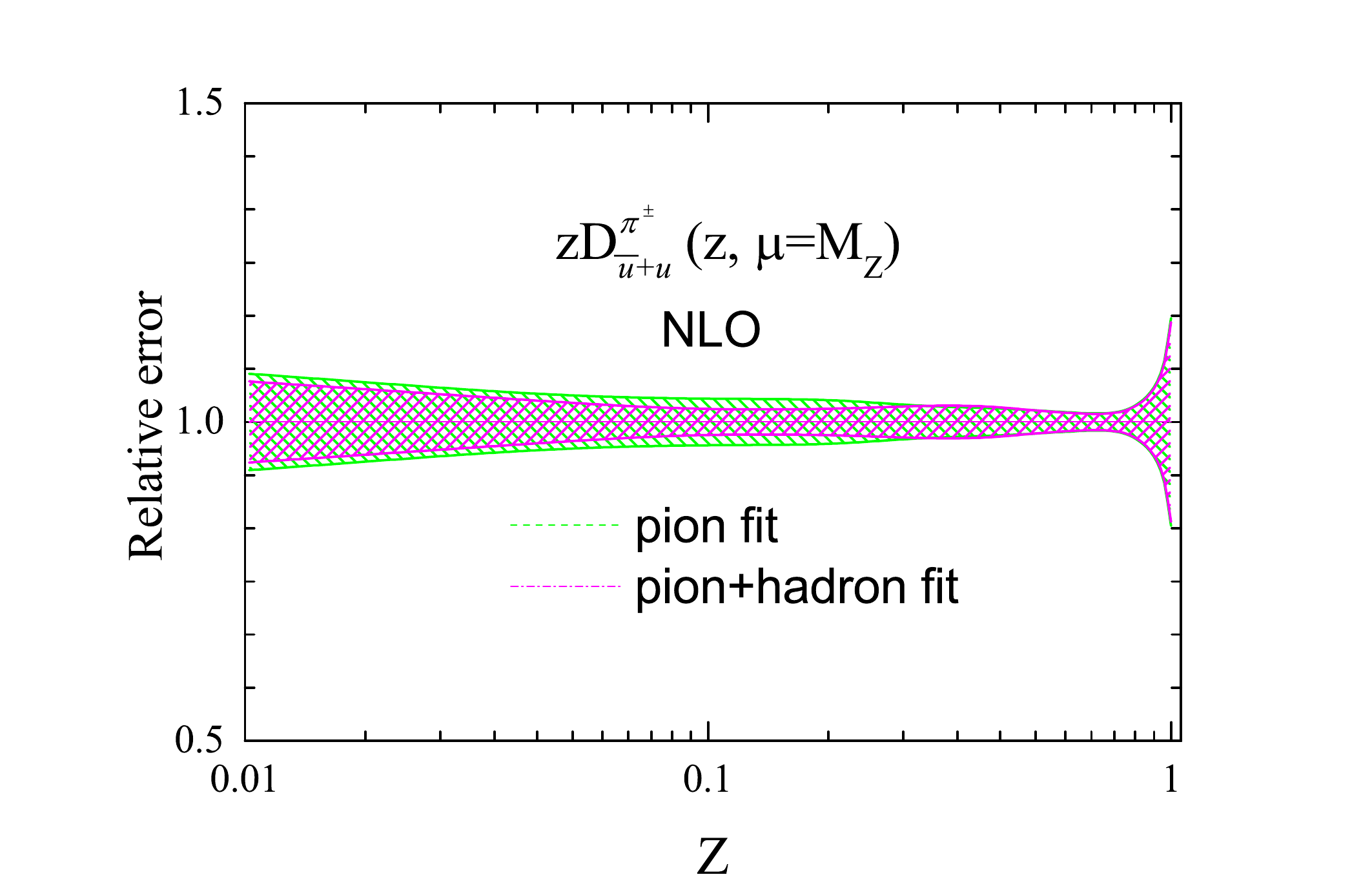}}  
	\resizebox{0.48\textwidth}{!}{\includegraphics{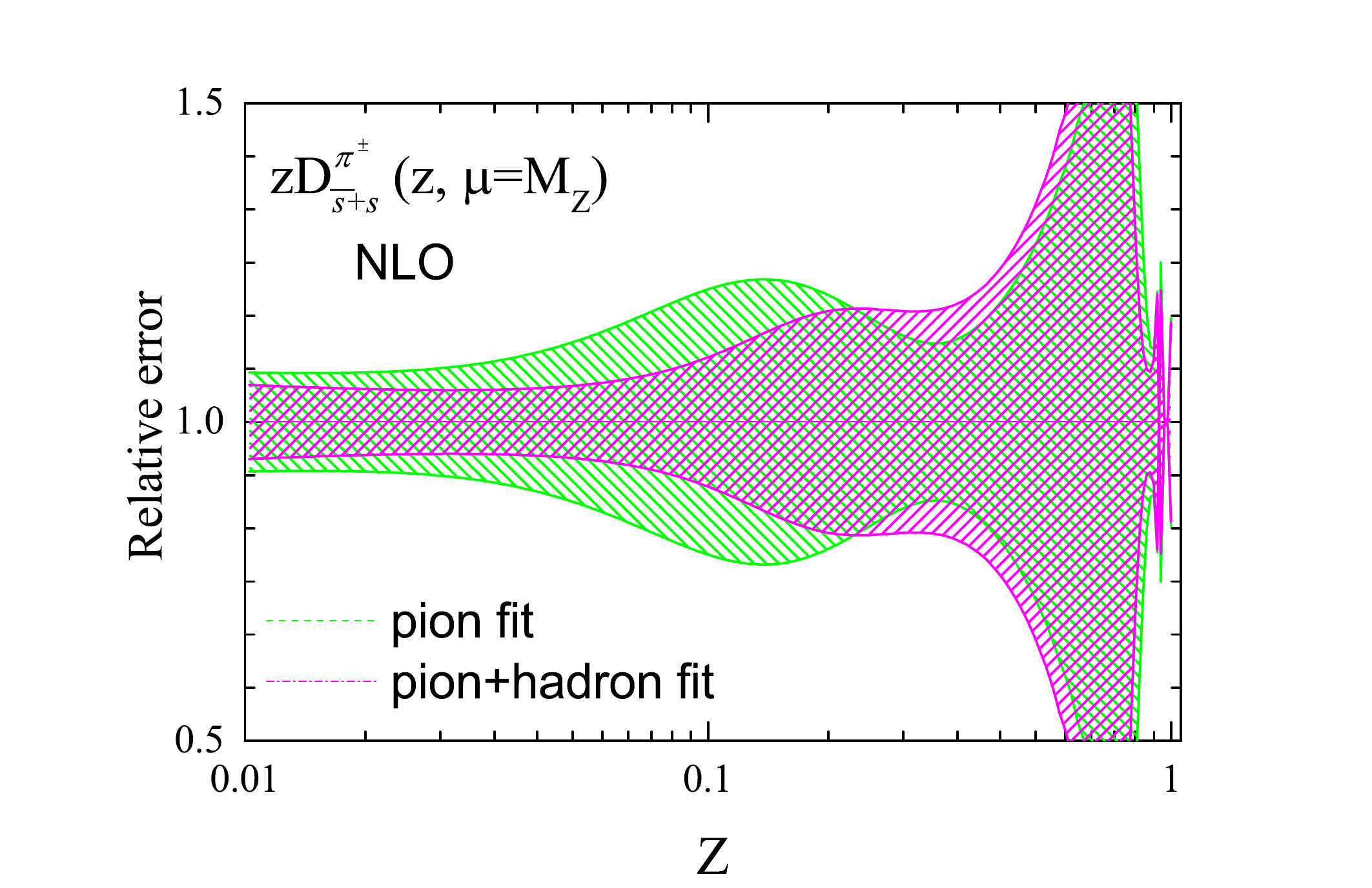}} 
	\resizebox{0.48\textwidth}{!}{\includegraphics{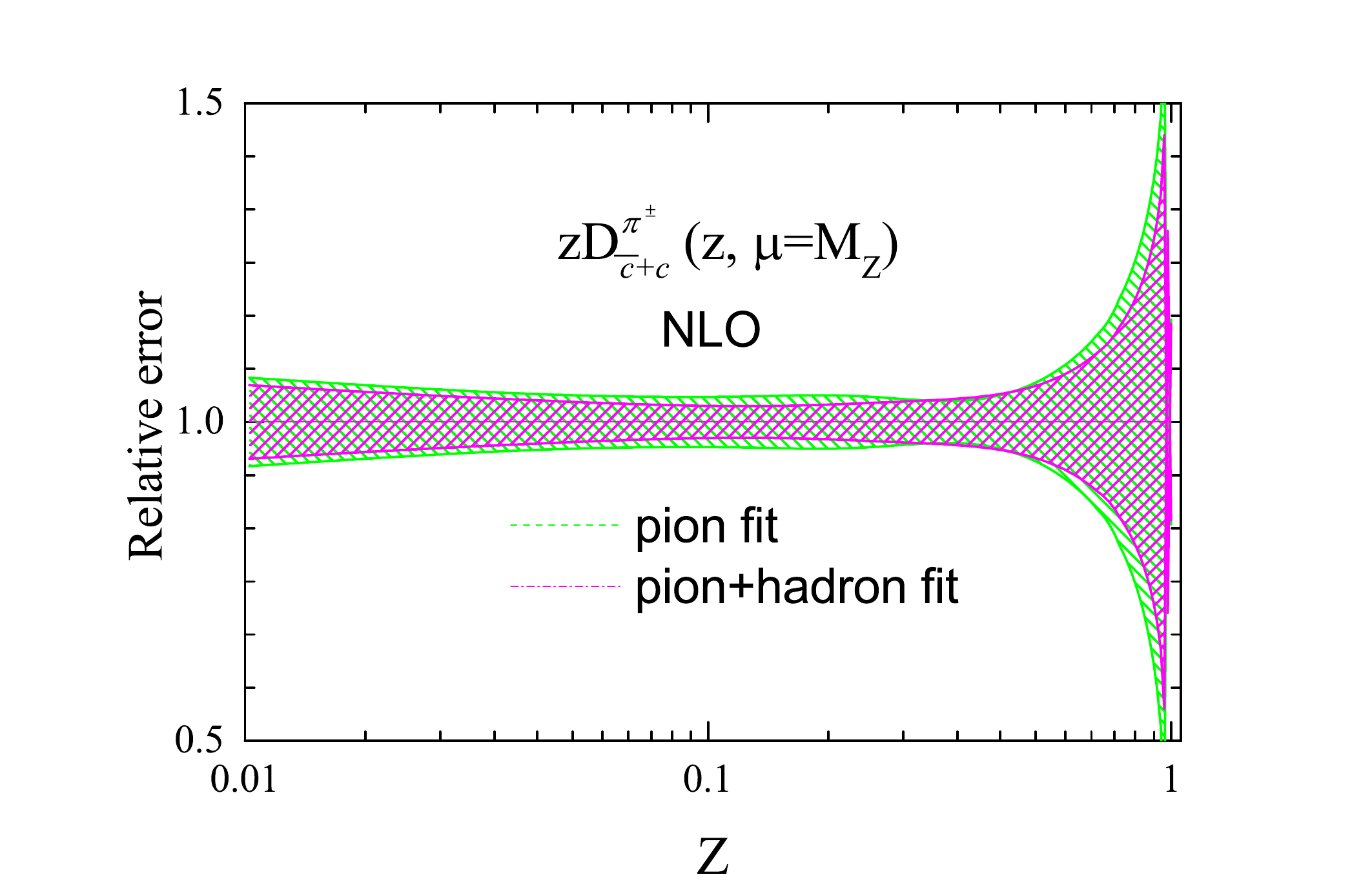}}  
	\resizebox{0.48\textwidth}{!}{\includegraphics{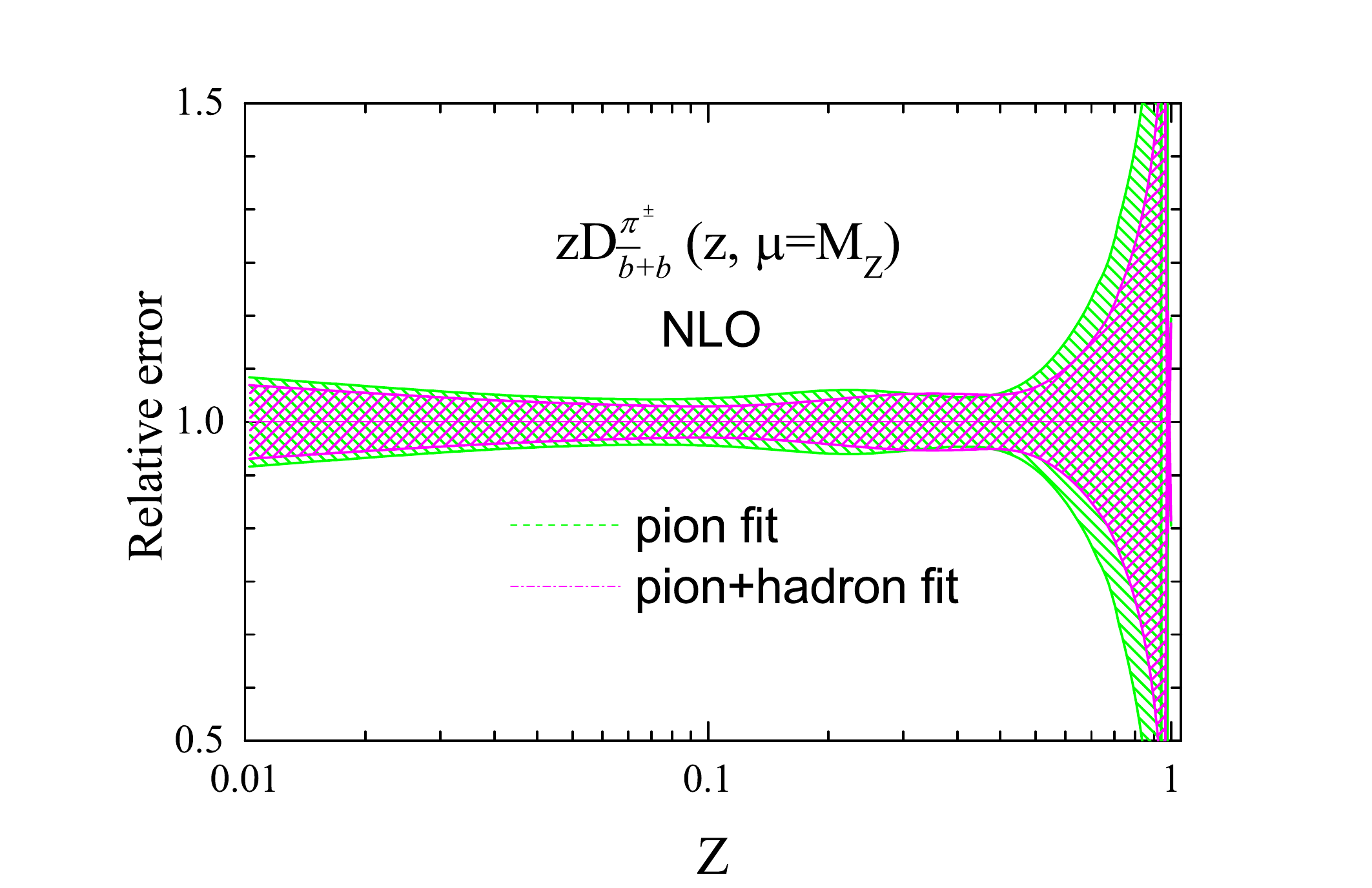}} 
	\resizebox{0.48\textwidth}{!}{\includegraphics{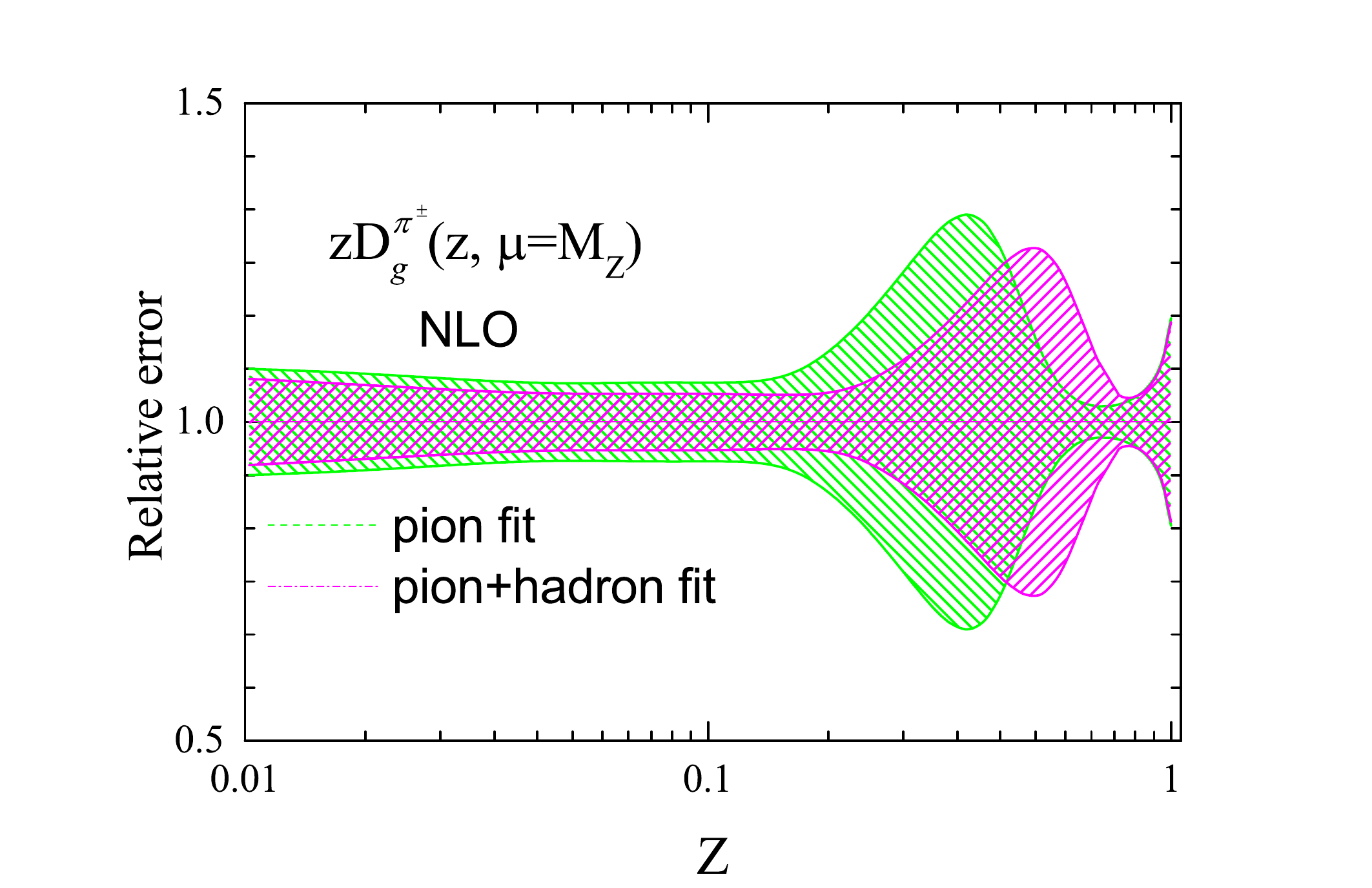}} 
	\begin{center}
		\caption{{\small Same as Fig.~\ref{fig:Ratioq5NLO} but at $Q=M_Z$. } \label{fig:RatioqMZNLO}}
	\end{center}
\end{figure*}

\begin{figure*}[htb]
	\vspace{0.50cm}
	\resizebox{0.48\textwidth}{!}{\includegraphics{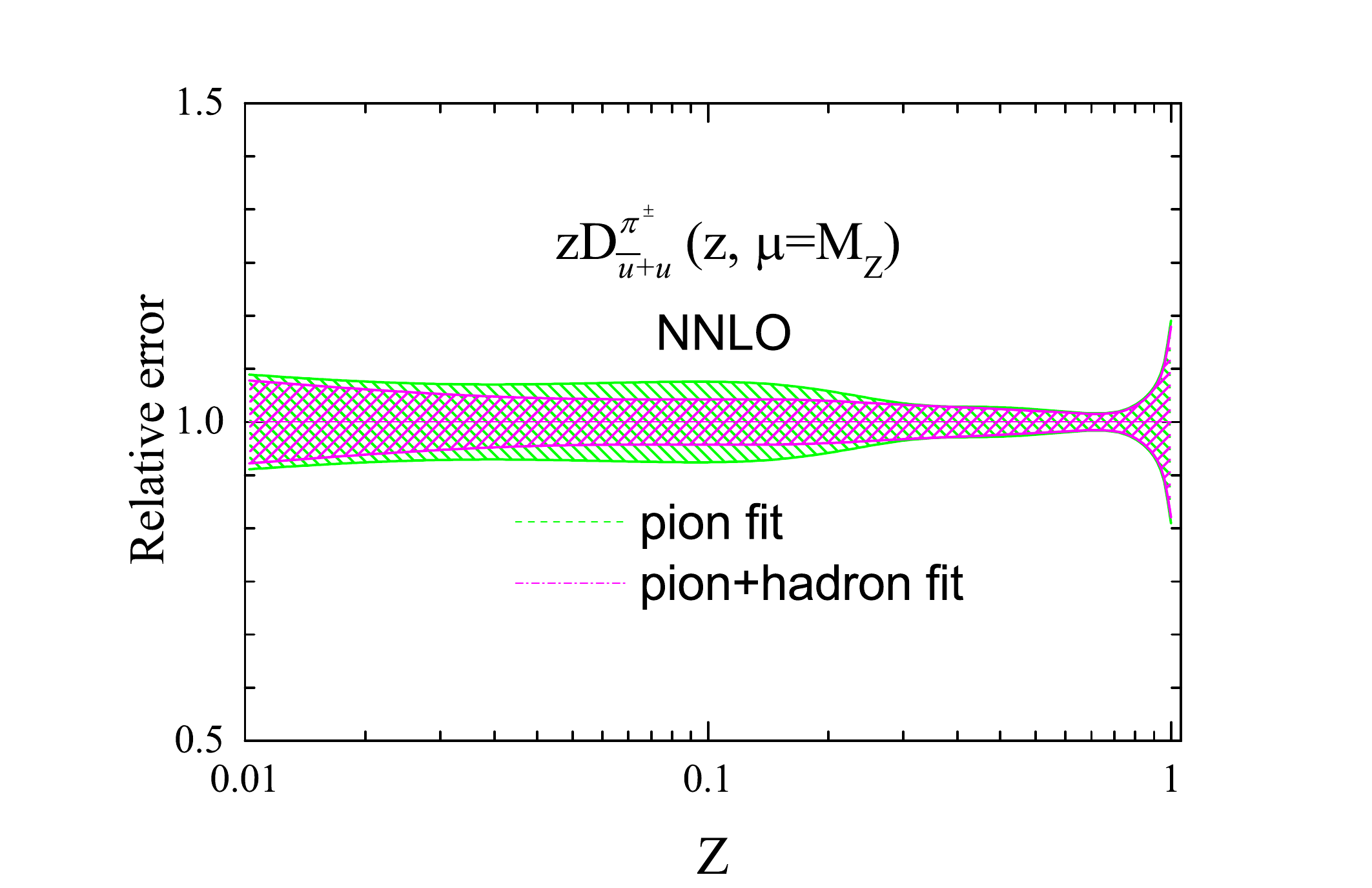}}  
	\resizebox{0.48\textwidth}{!}{\includegraphics{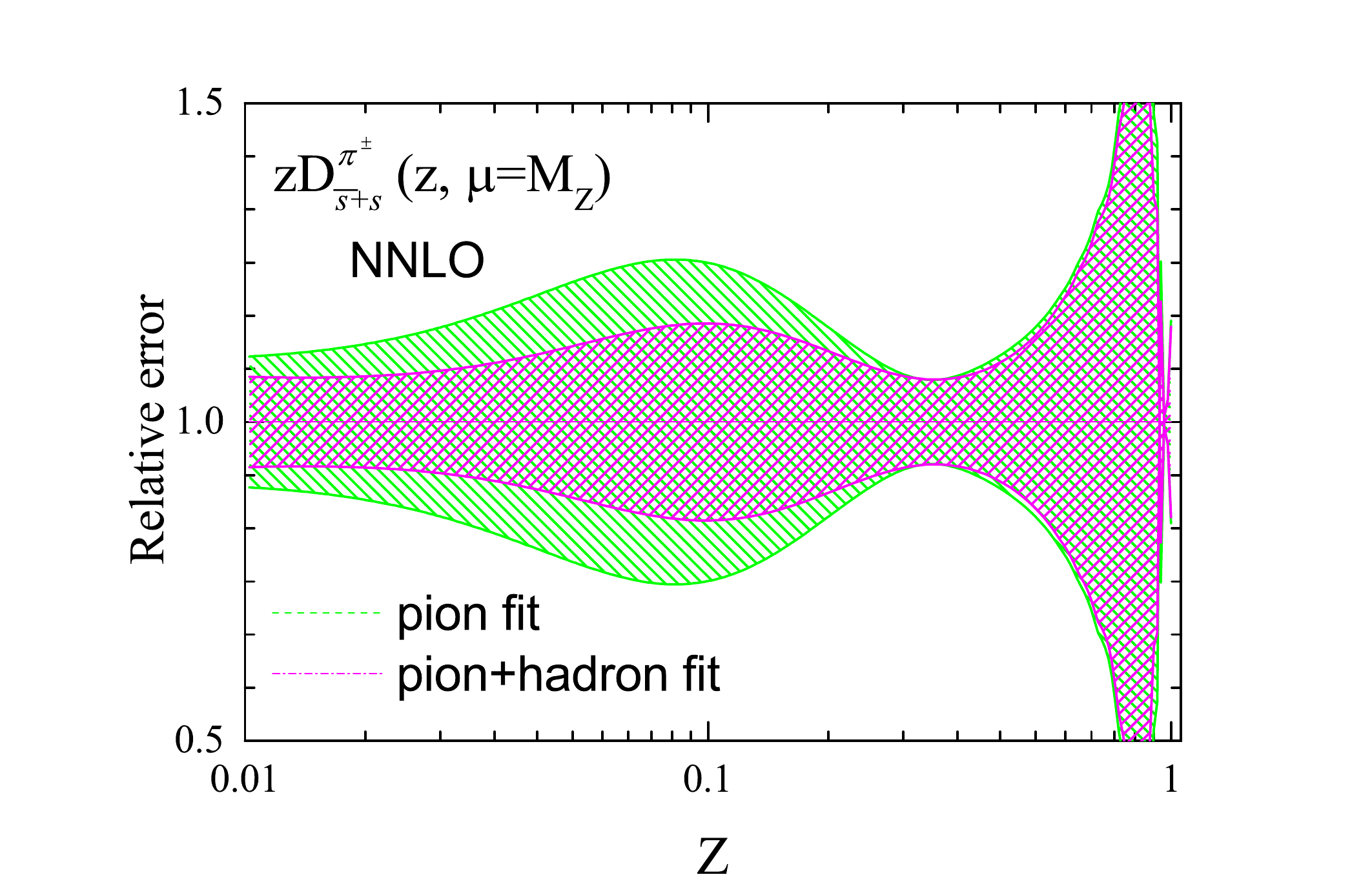}} 
	\resizebox{0.48\textwidth}{!}{\includegraphics{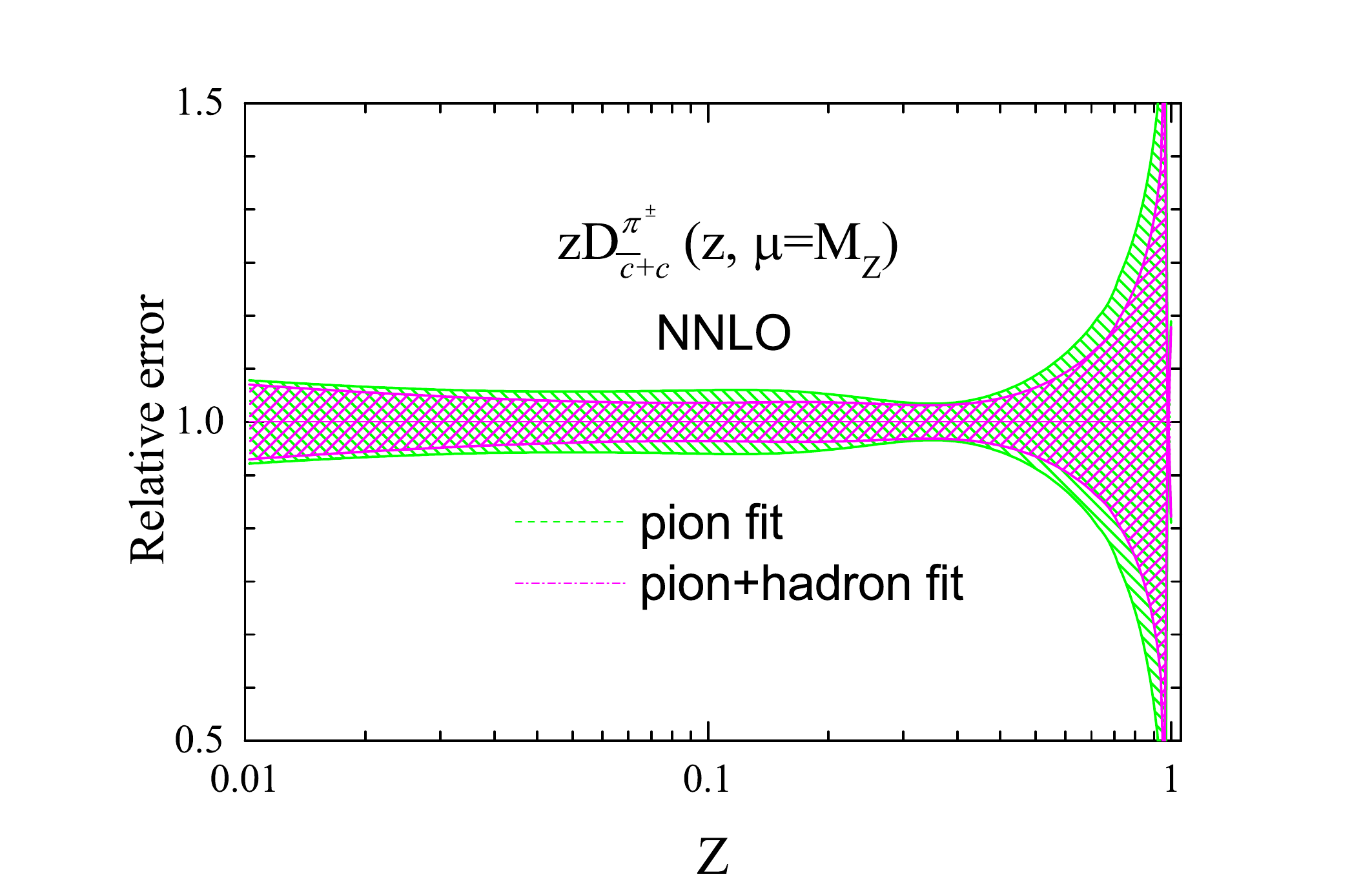}}  
	\resizebox{0.48\textwidth}{!}{\includegraphics{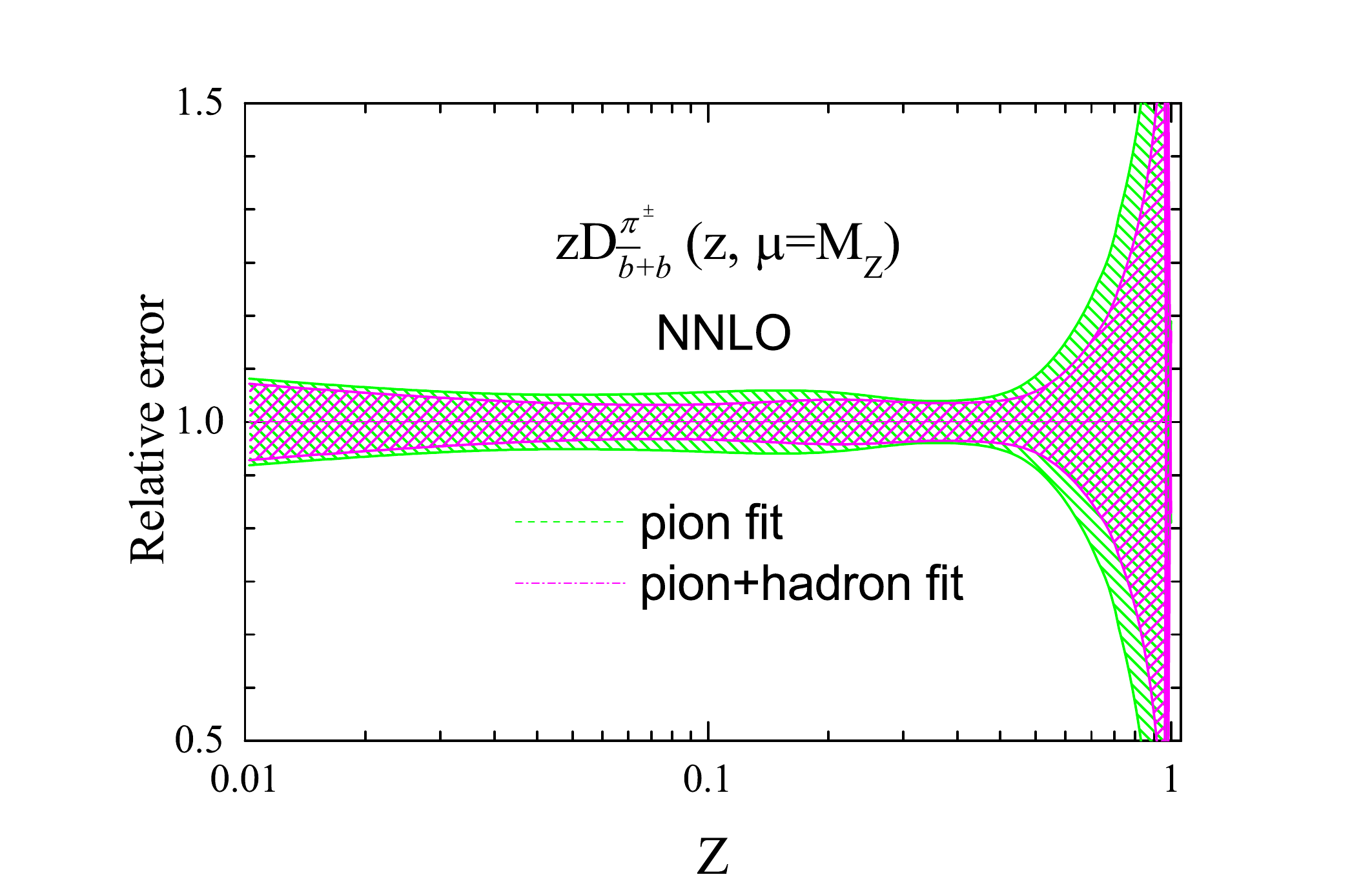}} 
	\resizebox{0.48\textwidth}{!}{\includegraphics{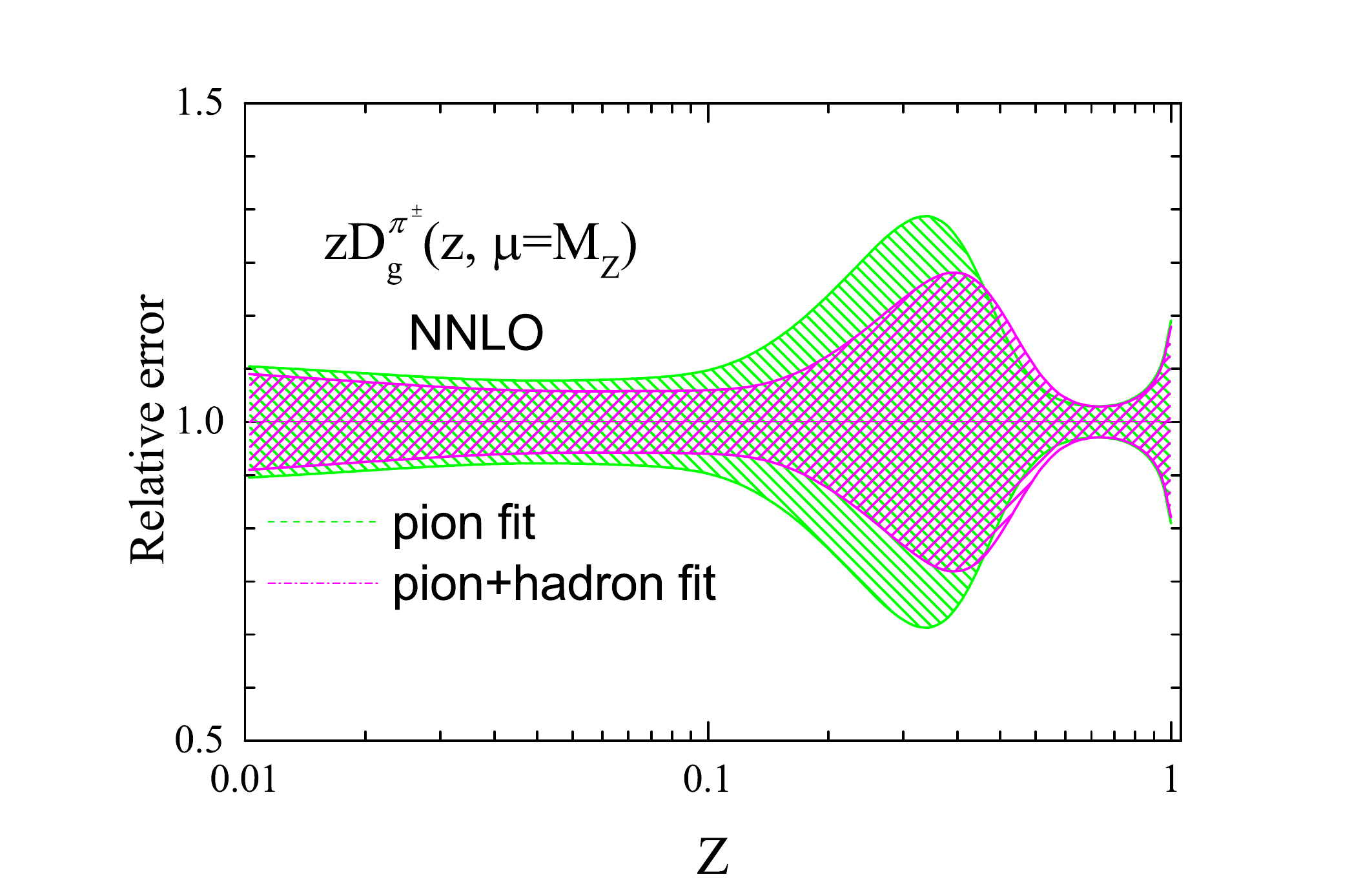}} 
	\begin{center}
		\caption{{\small Same as Fig.~\ref{fig:Ratioq5NLO} but for $Q=M_Z$ at NNLO.} \label{fig:RatioqMZNNLO}}
	\end{center}
\end{figure*}

\begin{figure*}[htb]
	\vspace{0.50cm}
	\resizebox{0.48\textwidth}{!}{\includegraphics{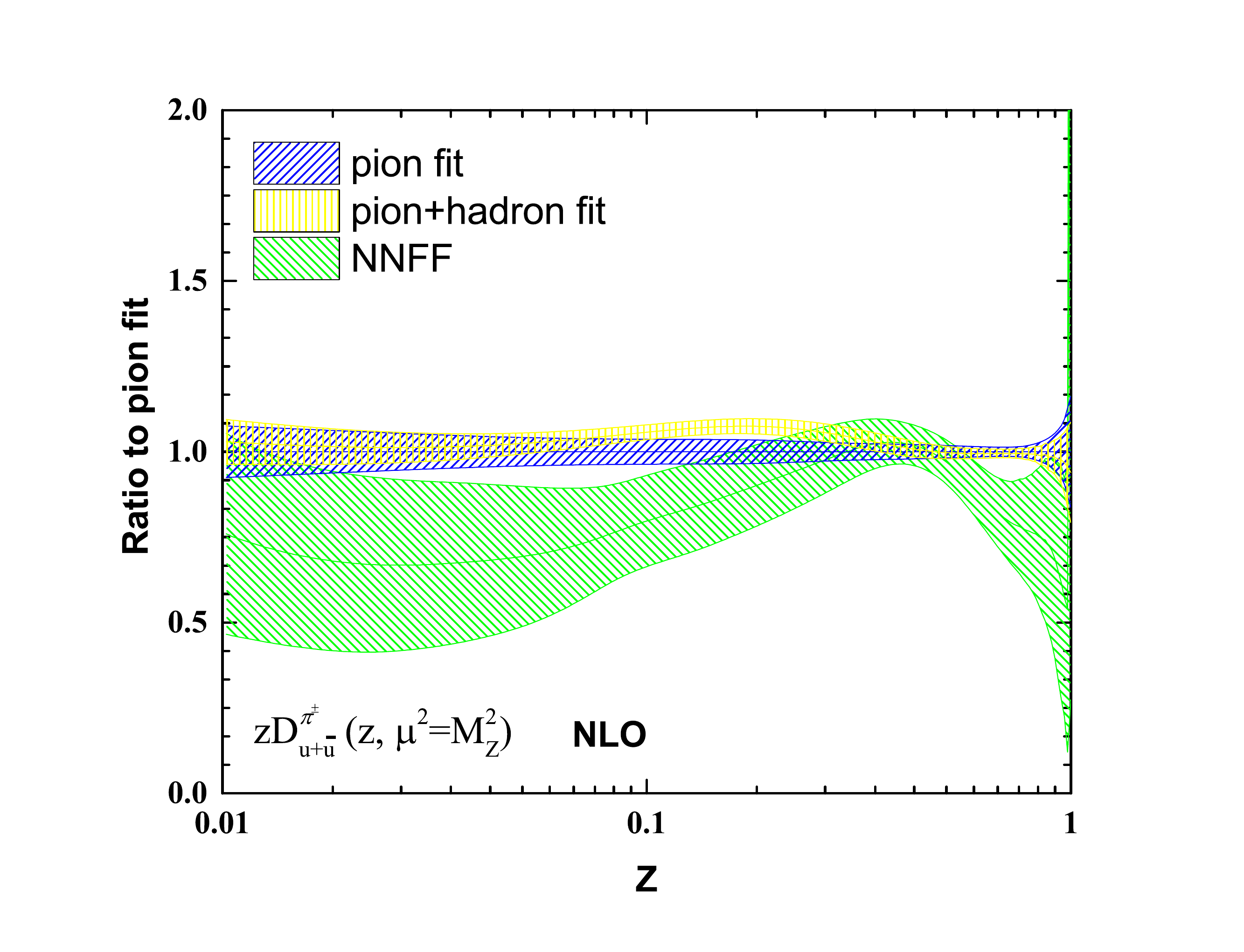}}  	
	\resizebox{0.48\textwidth}{!}{\includegraphics{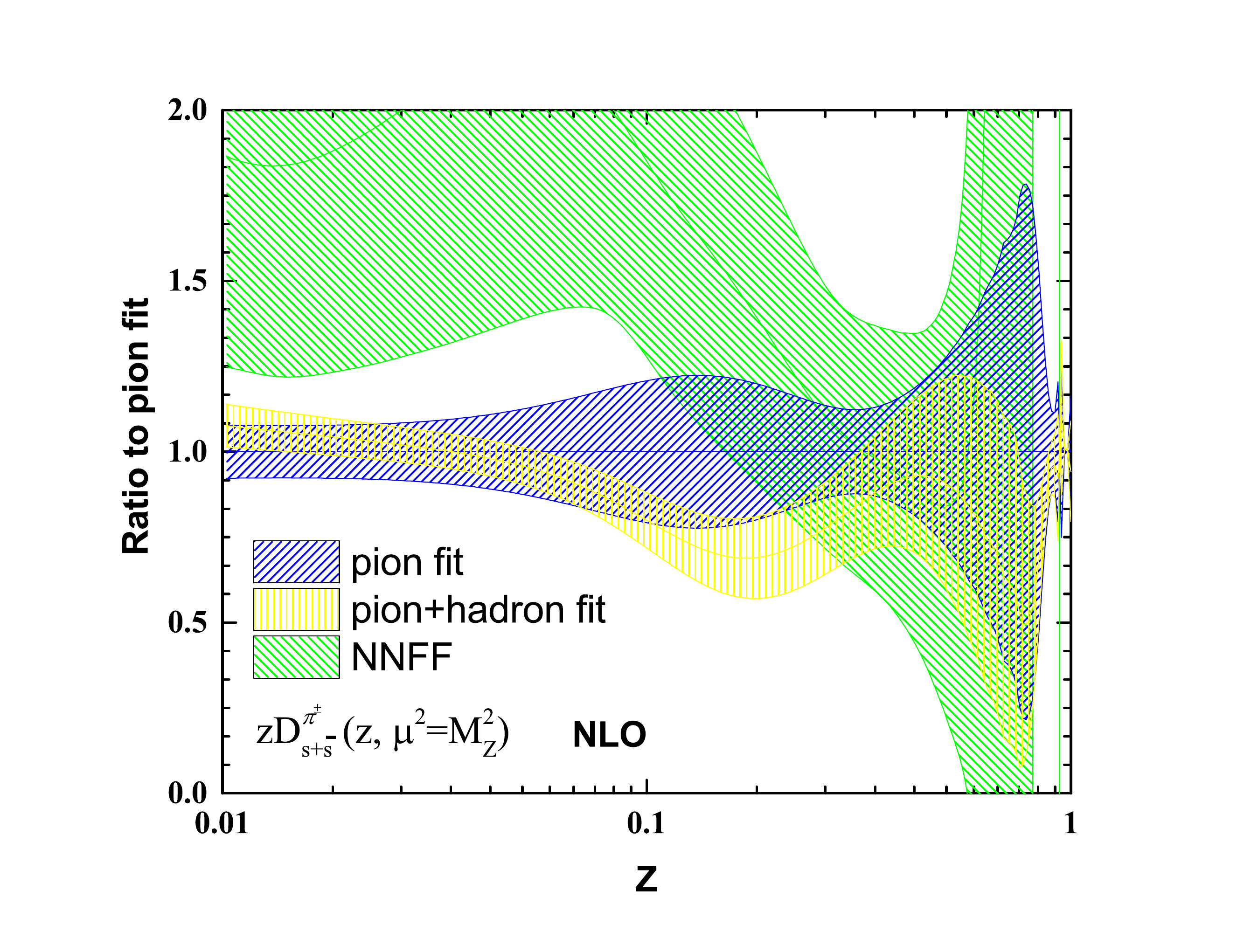}}  	
	\resizebox{0.48\textwidth}{!}{\includegraphics{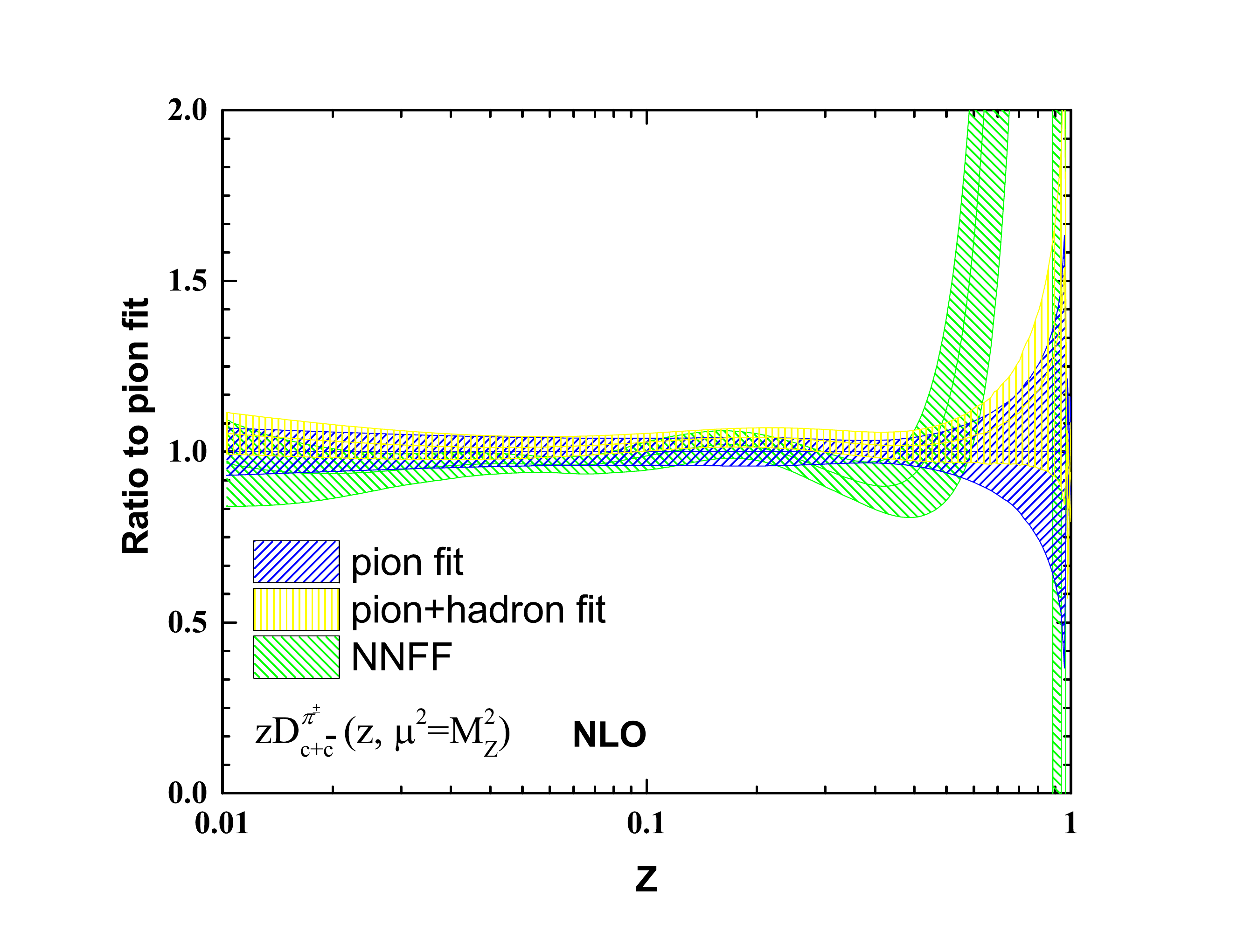}}  	
	\resizebox{0.48\textwidth}{!}{\includegraphics{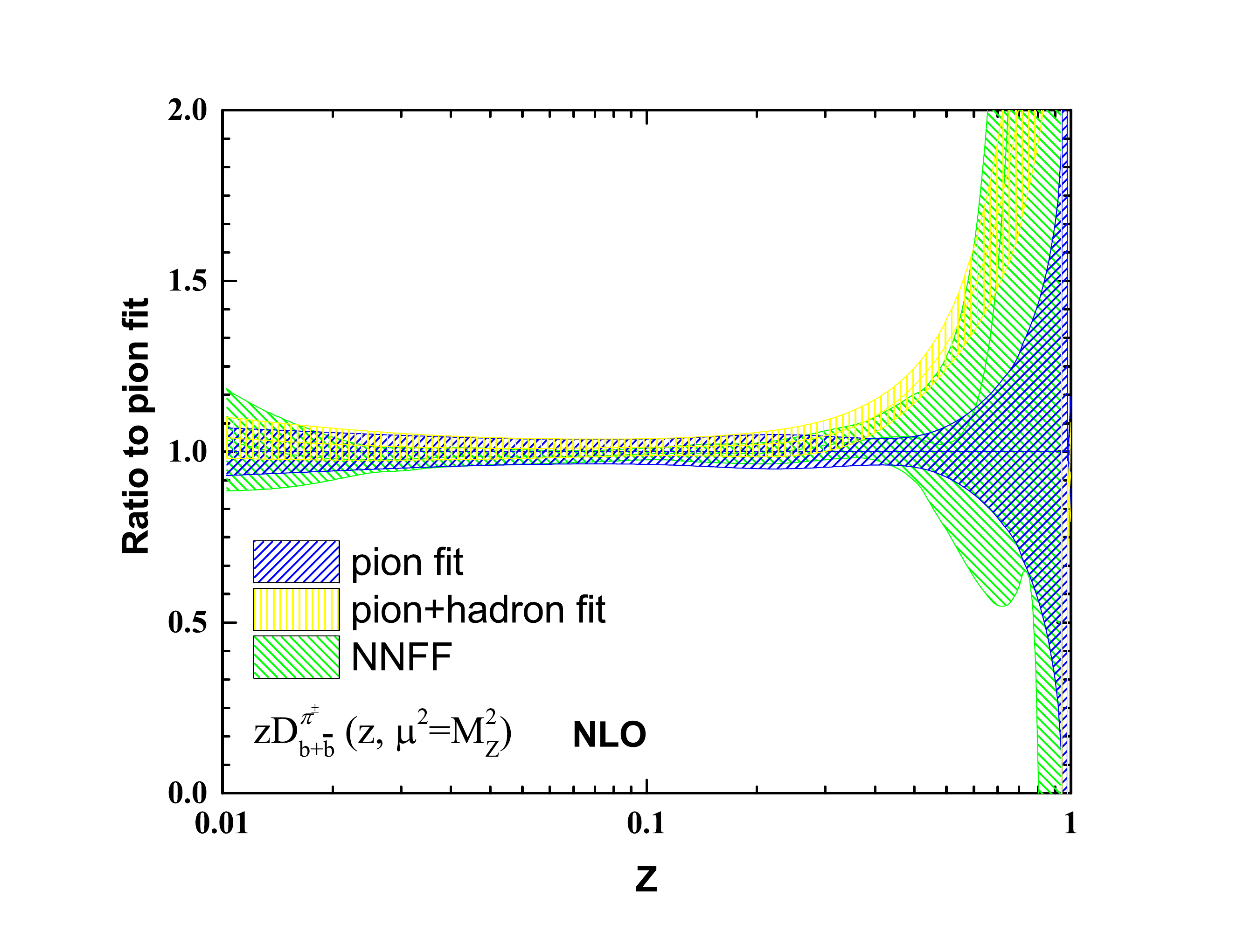}}  	
	\resizebox{0.48\textwidth}{!}{\includegraphics{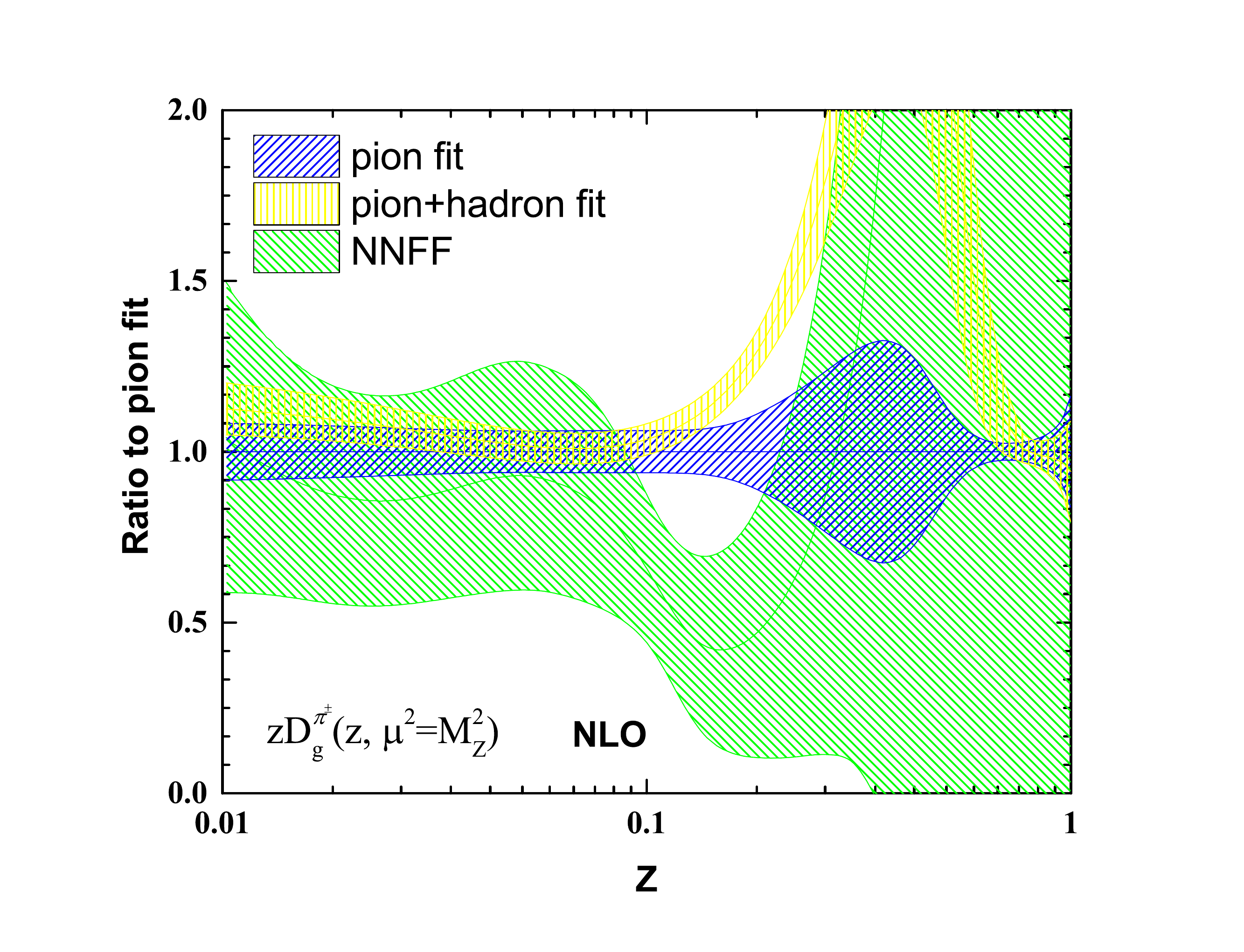}}  	
	\begin{center}
		\caption{{\small Comparison between the pion FFs ratios from the ``{\tt pion fit}",  ``{\tt pion+hadron fit}" and {\tt NNFF1.0} analyses to the pion FFs from ``pion"  analysis at NLO for $Q=M_Z$ . } \label{fig:RatioqMZNLOtopion}}
	\end{center}
\end{figure*}

\begin{figure*}[htb]
	\vspace{0.50cm}
	\resizebox{0.48\textwidth}{!}{\includegraphics{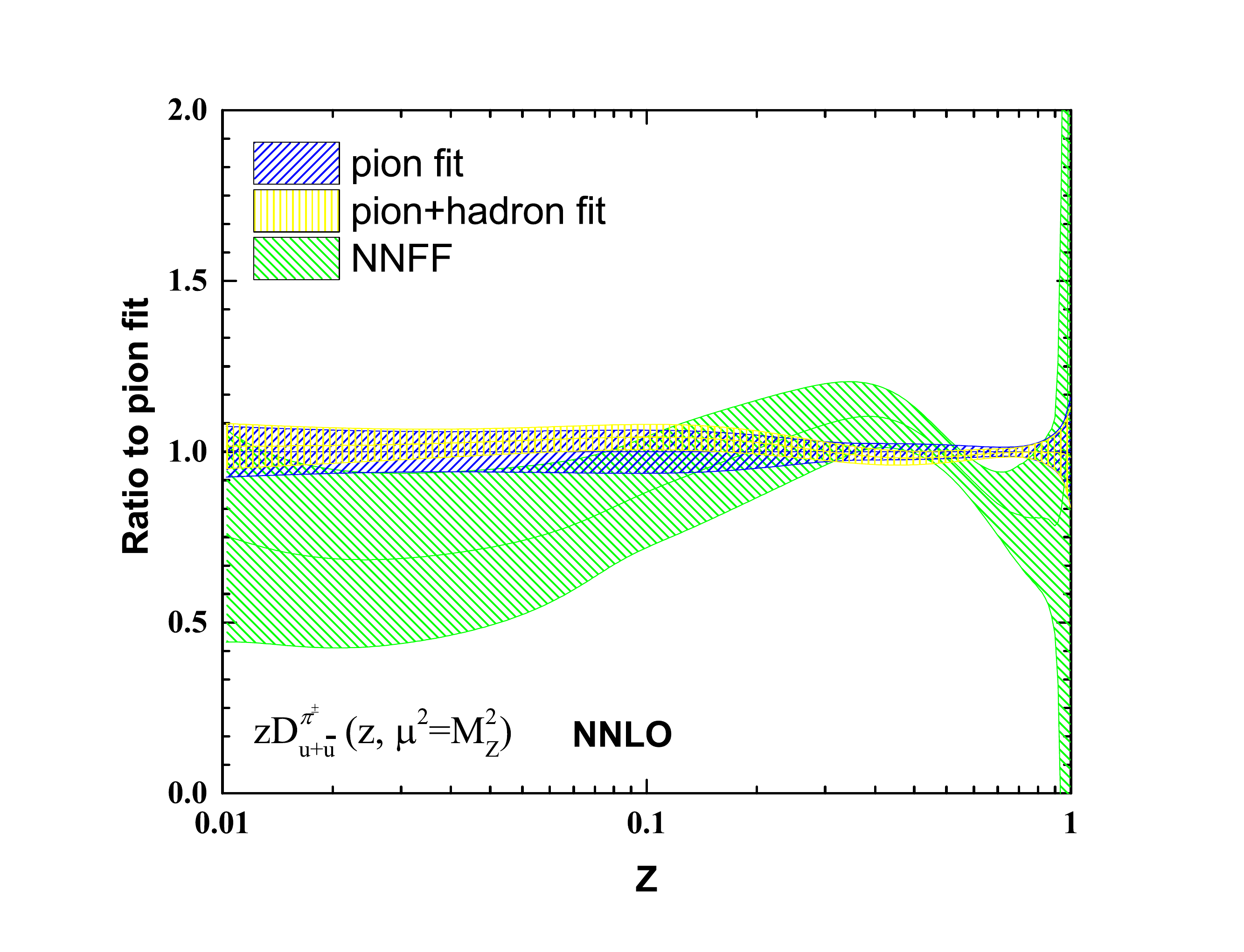}}  	
	\resizebox{0.48\textwidth}{!}{\includegraphics{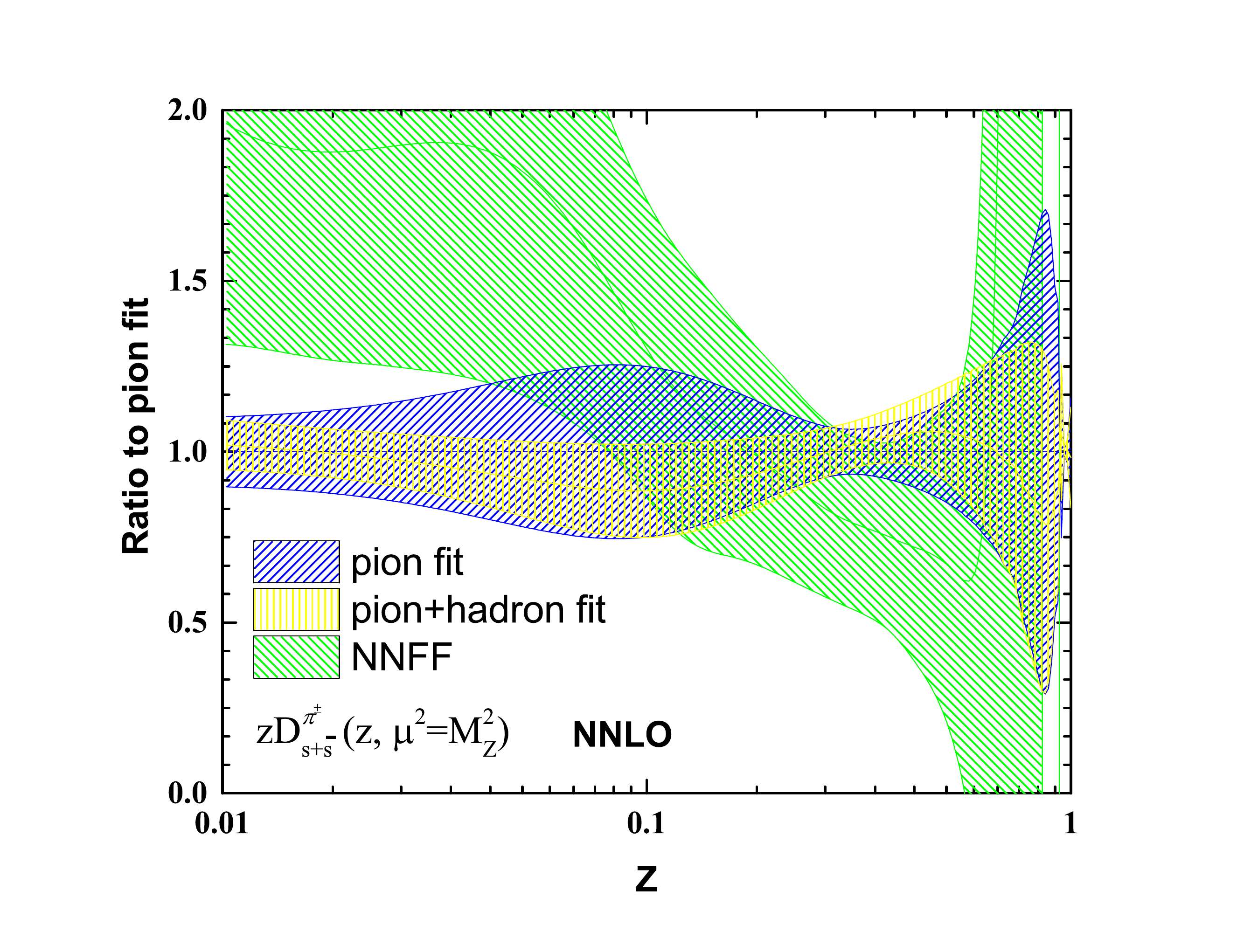}}  	
	\resizebox{0.48\textwidth}{!}{\includegraphics{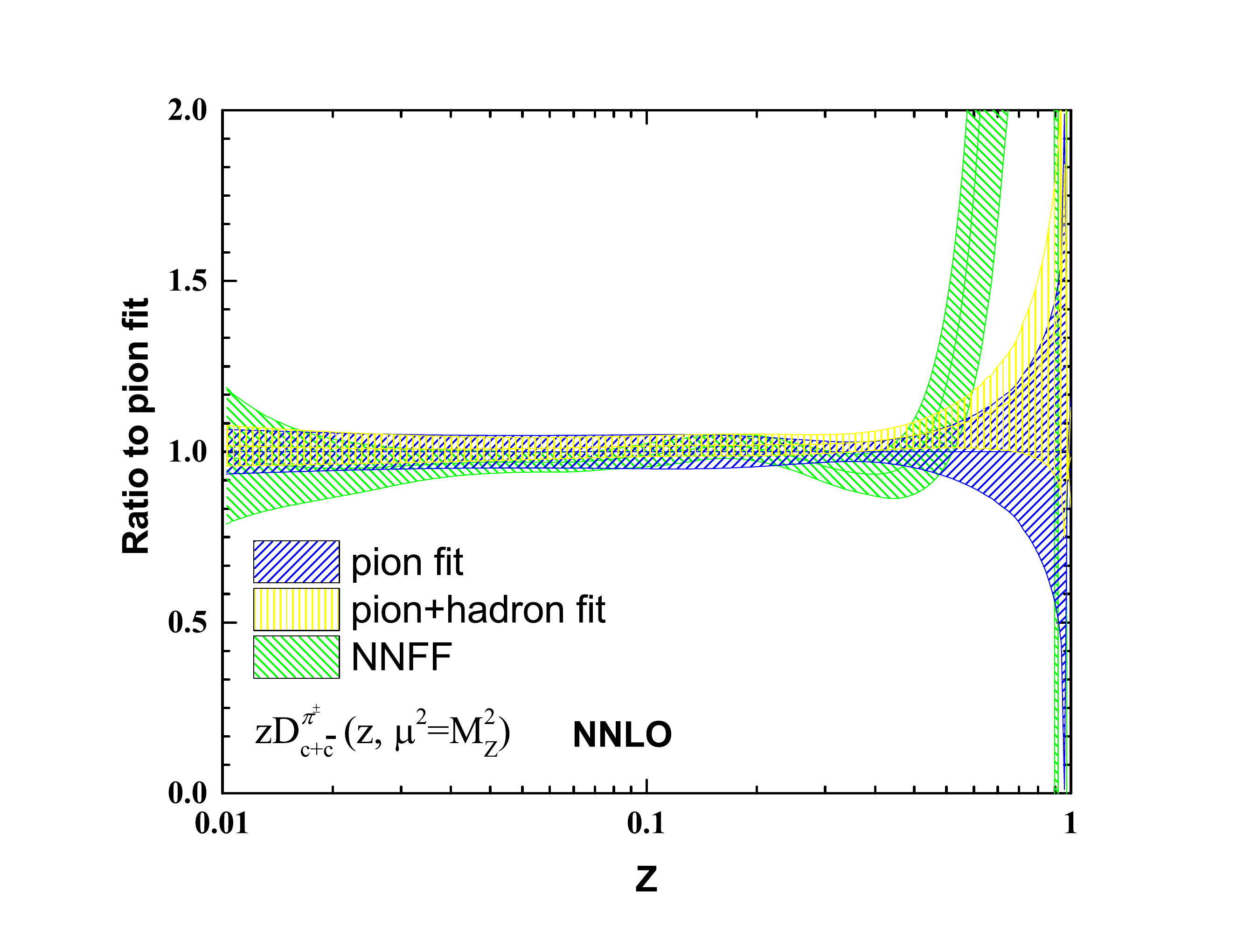}}  	
	\resizebox{0.48\textwidth}{!}{\includegraphics{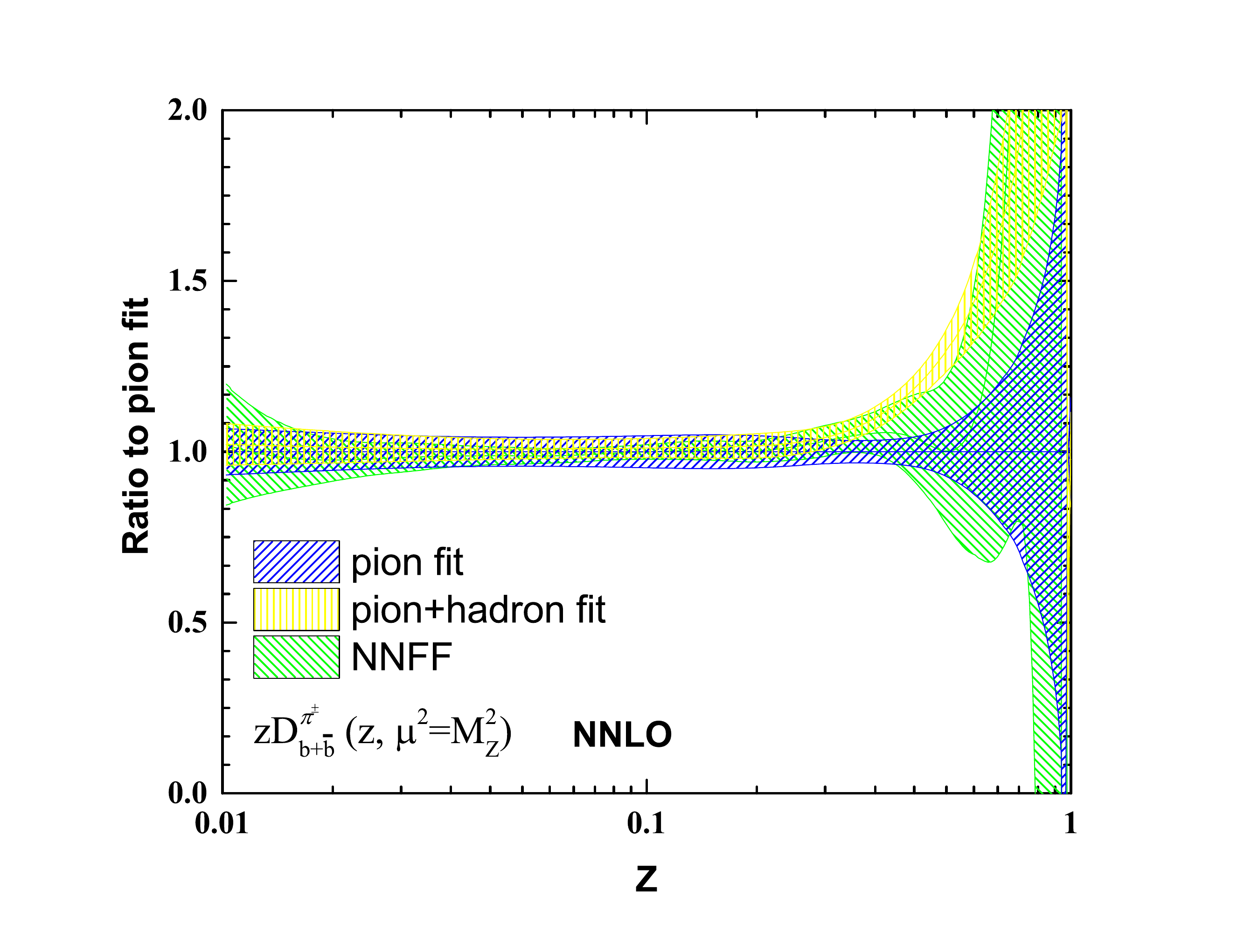}}  	
	\resizebox{0.48\textwidth}{!}{\includegraphics{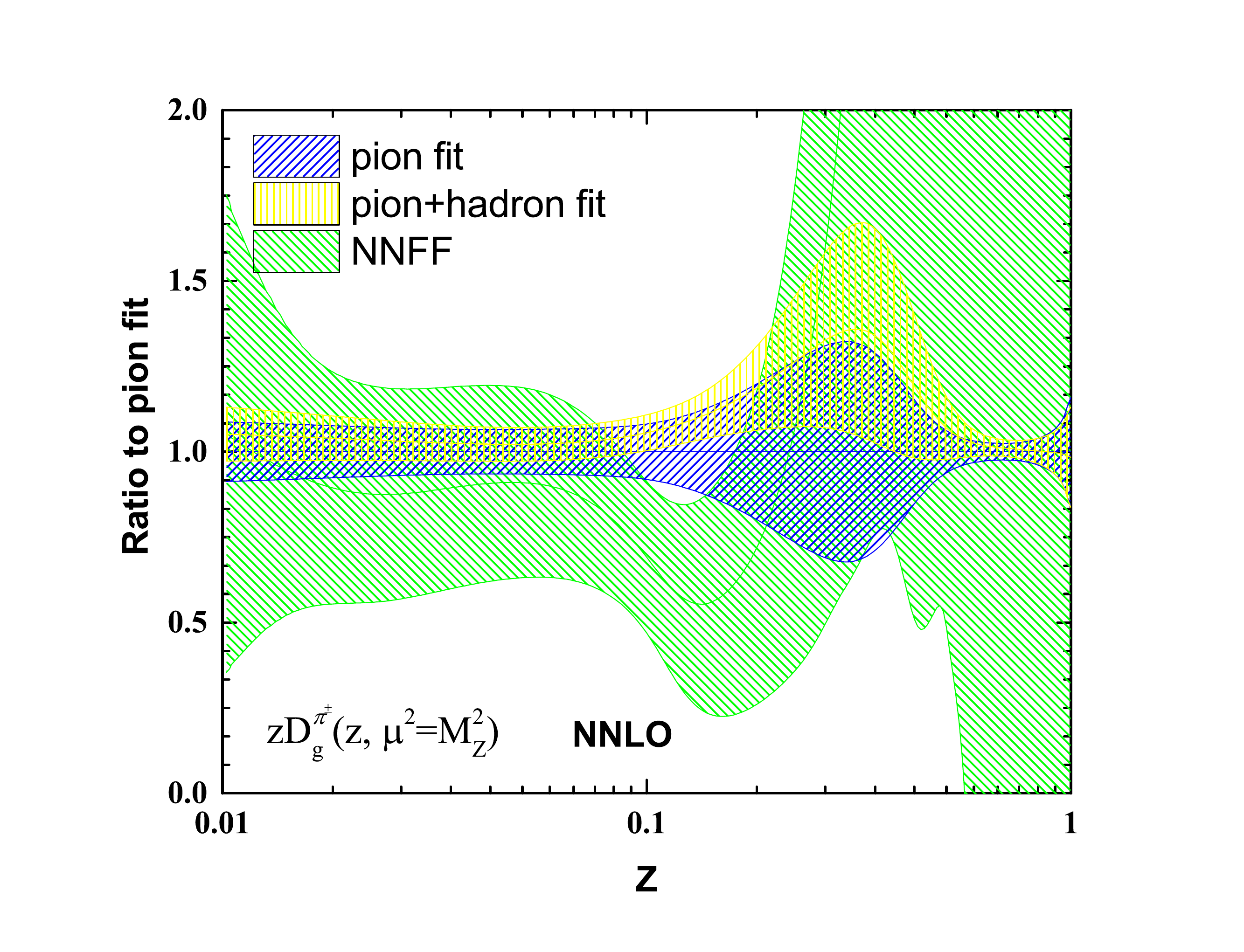}}  	
	\begin{center}
		\caption{{\small Same as Fig.~\ref{fig:RatioqMZNLOtopion} but at NNLO. } \label{fig:RatioqMZNNLOtopion}}
	\end{center}
\end{figure*}

Fig.~\ref{fig:Ratioq5NNLO} shows the same results as Fig.~\ref{fig:Ratioq5NLO}, but  this time for our NNLO analysis. One can clearly conclude that the inclusion of the unidentified light charged hadron data in the pion FFs analysis at NNLO accuracy can also lead to a smaller relative uncertainty for all flavors. Note that, compared with the NLO results, the relative uncertainty of $s+\bar s$ FF from ``{\tt pion+hadron fit}" analysis has now remarkably decreased at lower $z$ values rather than its distribution from ``{\tt pion fit}" analysis. Overall, the results obtained indicate that by performing a simultaneous analysis of pion and unidentified light charged hadron data, a pion FFs set with more acceptable uncertainties can be obtained at both NLO and NNLO accuracies. 

To study the effects of the evolution and also evaluate the results at a given higher energy, we recalculate the predictions of Figs.~\ref{fig:Ratioq5NLO} and~\ref{fig:Ratioq5NNLO}, but this time for $Q=M_z$. The results obtained have been shown in Figs.~\ref{fig:RatioqMZNLO} and~\ref{fig:RatioqMZNNLO} at NLO and NNLO, respectively. The reduction in the relative uncertainty of all flavors after the inclusion of the unidentified light charged hadron data in the analysis is clearly seen from these figures. Note that the shift observed in the relative uncertainty of $s+\bar s$ and gluon FFs from ``{\tt pion+hadron fit}" analysis compared with the ``{\tt pion fit}" analysis at NLO (see Fig.~\ref{fig:RatioqMZNLO}) is due to the considerable change in the central values of these distributions after the inclusion of the hadron data.

Another way for comparing the results of two aforementioned analyses is using the ratio plots in which any change in the central values of the distribution can be also investigated, in addition to their uncertainties. Fig.~\ref{fig:RatioqMZNLOtopion} shows a comparison between the ratios of pion FFs obtained from the ``{\tt pion+hadron fit}" analysis (yellow band) and also {\tt NNFF1.0}~\cite{Bertone:2017tyb} (green band) to those obtained from the ``{\tt pion fit}" analysis (blue band) at $Q=M_z$ and NLO. According to the results obtained, one can sees that the uncertainties of all flavor distributions have been decreased by inclusion the unidentified light charged hadron data in the analysis compared with the ``{\tt pion fit}" analysis. Overall, our FFs whether from the ``{\tt pion fit}" analysis or ``{\tt pion+hadron fit}" one, have less uncertainties than the {\tt NNFF1.0} results, especially for the case of up, strange and gluon distributions.

Let us focus on each flavor separately to discuss about the changes in more details. For the case of $u + \bar u$ FF, no significant change can be seen between the ``{\tt pion fit}" and ``{\tt pion+hadron fit}" analyses. However, both of these analyses have different results than the $u+\bar u$ FF of {\tt NNFF1.0}, almost for all values of $z$. Actually, the difference is more significant at lower values of $z$ and reaches even to 30\%. The second panel of Fig.~\ref{fig:RatioqMZNLOtopion} shows that the inclusion of hadron data in the analysis of pion FFs at NLO can put further constraints on $s+\bar s$ FF, especially at medium to small $z$ regions, so that the uncertainty is remarkably reduced. Moreover, it decreases the $s+\bar s$ distribution in magnitude at medium and large values of $z$. It should be noted that our results for the $s+\bar s$ FF are very different to {\tt NNFF1.0} result and have smaller magnitude up to 100\% at smaller $z$ values.
For the case of $c+\bar c$ and $b+\bar b$ FFs, all three analyses have almost same results both in magnitude and uncertainties at medium to small values of $z$, but differ at larger values. To be more precise, the $c+\bar c$ FF of ``{\tt pion fit}" and ``{\tt pion+hadron fit}" analyses are similar even at large values of $z$, but the {\tt NNFF1.0} result is grows rapidly in this region. In contrast, the $b+\bar b$ FF of ``{\tt pion+hadron fit}" analysis behaves more similar to the {\tt NNFF1.0} and grows rapidly at large $z$ values compared with the ``{\tt pion fit}" analysis. Overall, one can conclude that the inclusion of the hadron data in the analysis does not affect the $c+\bar c$ FF, but can change the $b+\bar b$ FF at large values of $z$. The last panel of Fig.~\ref{fig:RatioqMZNLOtopion} shown again the immense impact of the unidentified light charged hadron data on the gluon FF of pion, especially at medium values of $z$. As can be seen, in addition to the significant reduction of the gluon FF uncertainty, its central value has changed considerably at around $z=0.4$ and become more consistent with the {\tt NNFF1.0} result at this region. However, there are still some differences at $0.1 \lesssim z \lesssim 0.8$, though all three analyses have almost same results at small $z$ values. Another important point should be noted is the very less uncertainty of our results compared with the {\tt NNFF1.0} one, in particular at large $z$ regions which can be attributed to the low flexibility of our parameterization for the gluon FF. 

Fig.~\ref{fig:RatioqMZNNLOtopion} shows the same results as Fig.~\ref{fig:RatioqMZNLOtopion}, but at NNLO accuracy. Overall, the interpretation of results obtained for each flavor distribution is similar to NLO case, with the difference that now the discrepancy observed between the $s+\bar s$ and also gluon FFs from ``{\tt pion fit}" and ``{\tt pion+hadron fit}" analyses at medium $z$ regions is more moderate than before. For example, the difference between the gluon FFs obtained from these two analyses at $z \simeq 0.4$ is less than 50\% according to the last panel of Fig.~\ref{fig:RatioqMZNNLOtopion}, while it is more than 100\% at NLO (see Fig.~\ref{fig:RatioqMZNLOtopion}). Another point should be noted is that the $u + \bar u$ and $c + \bar c$ FFs remain still unchanged after the inclusion of the unidentified light charged hadron data in the analysis, and the $b + \bar b$ FF is rapidly grown at large $z$ values just similar to NLO case.

\begin{figure*}[htb]
	\vspace{0.50cm}
	\resizebox{0.48\textwidth}{!}{\includegraphics{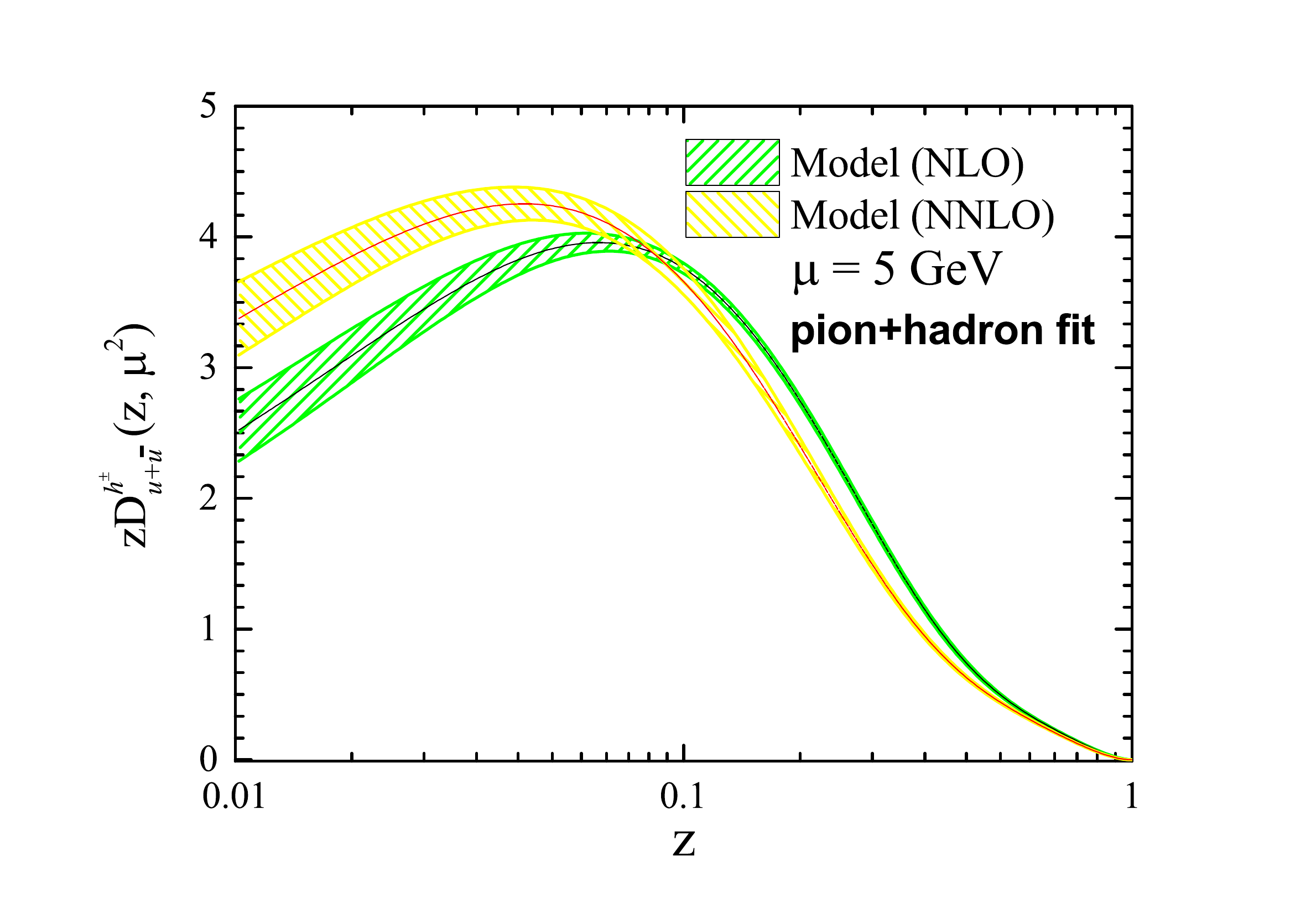}}  	
	\resizebox{0.48\textwidth}{!}{\includegraphics{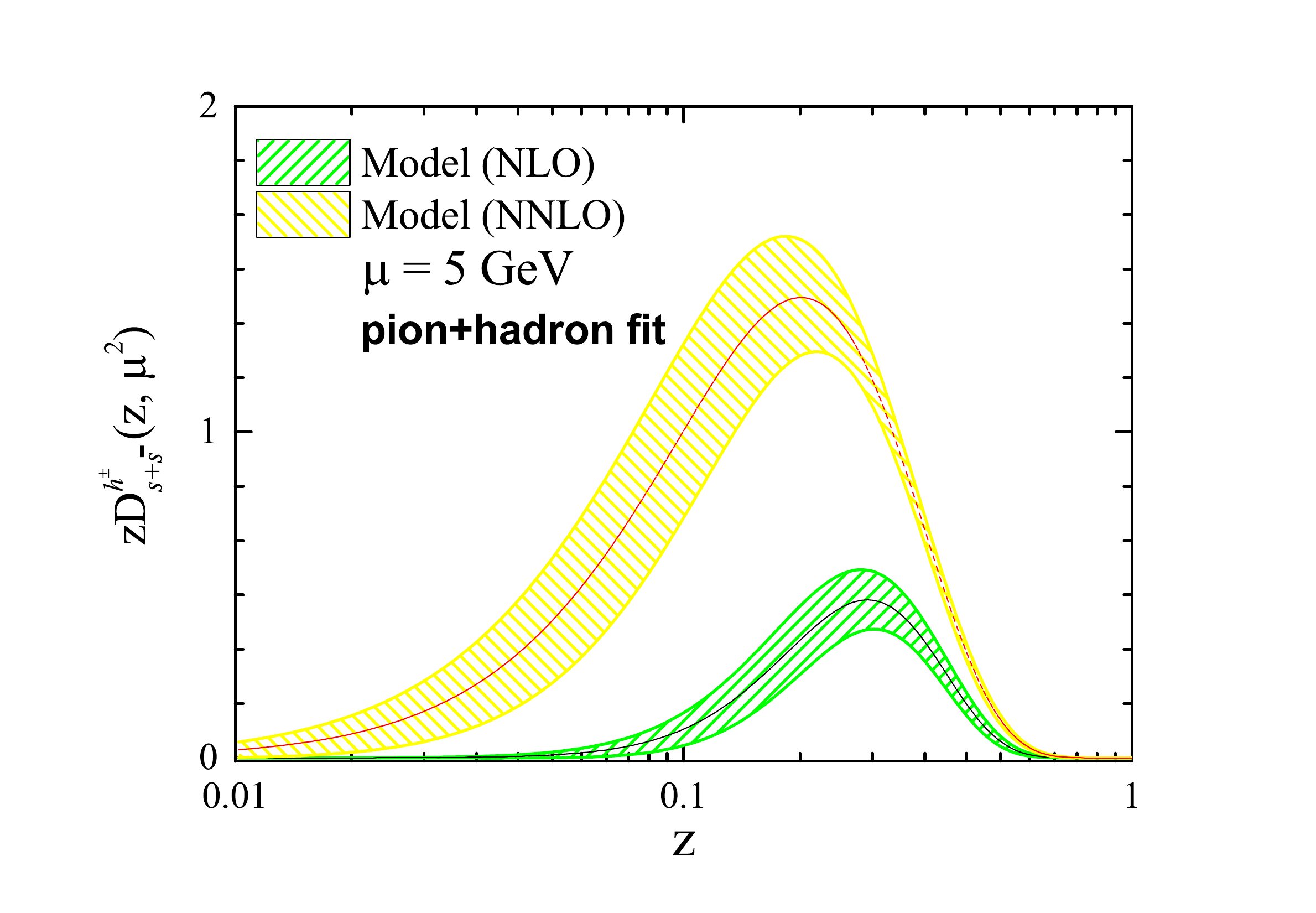}}  	
	\resizebox{0.48\textwidth}{!}{\includegraphics{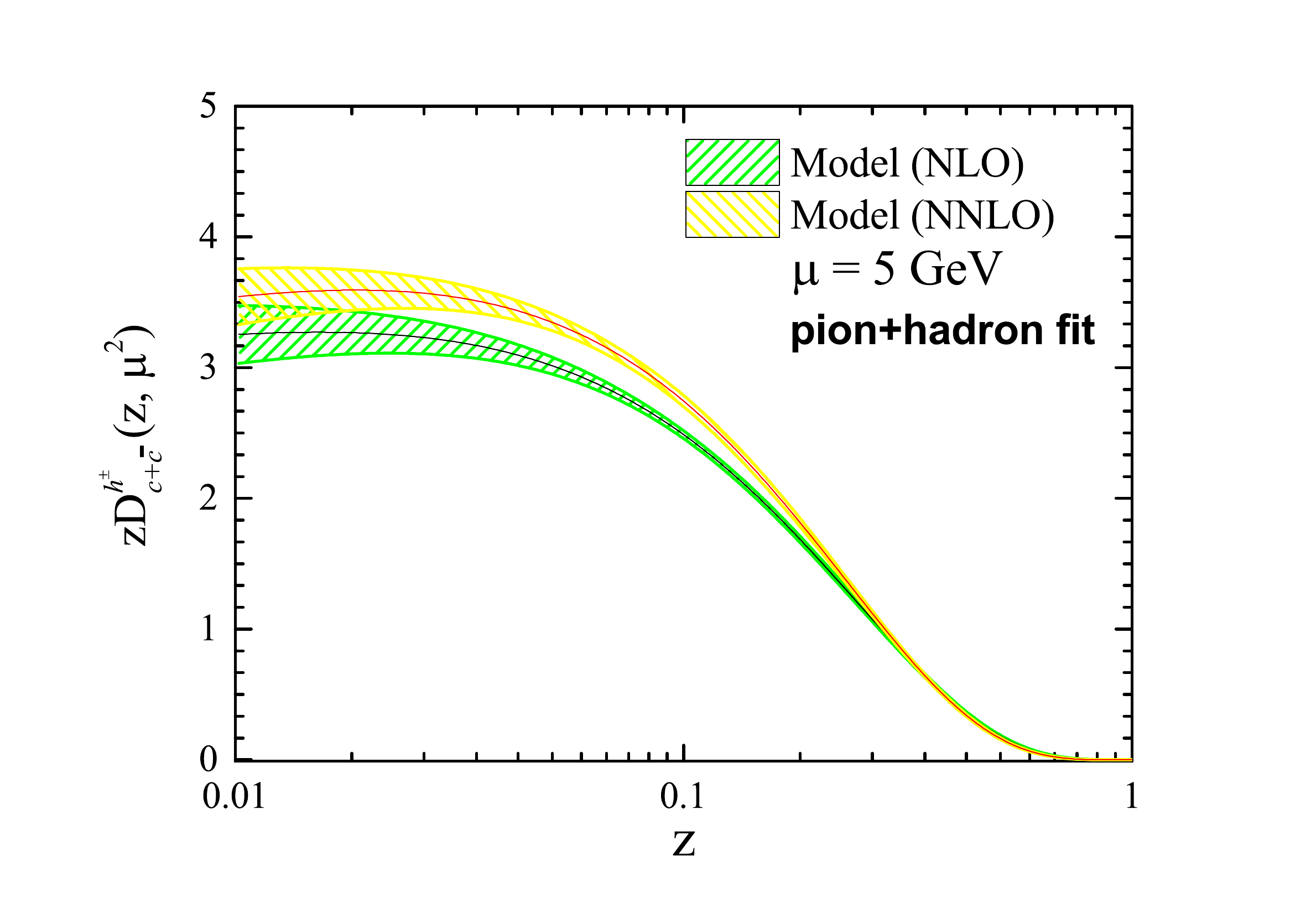}}  	
	\resizebox{0.48\textwidth}{!}{\includegraphics{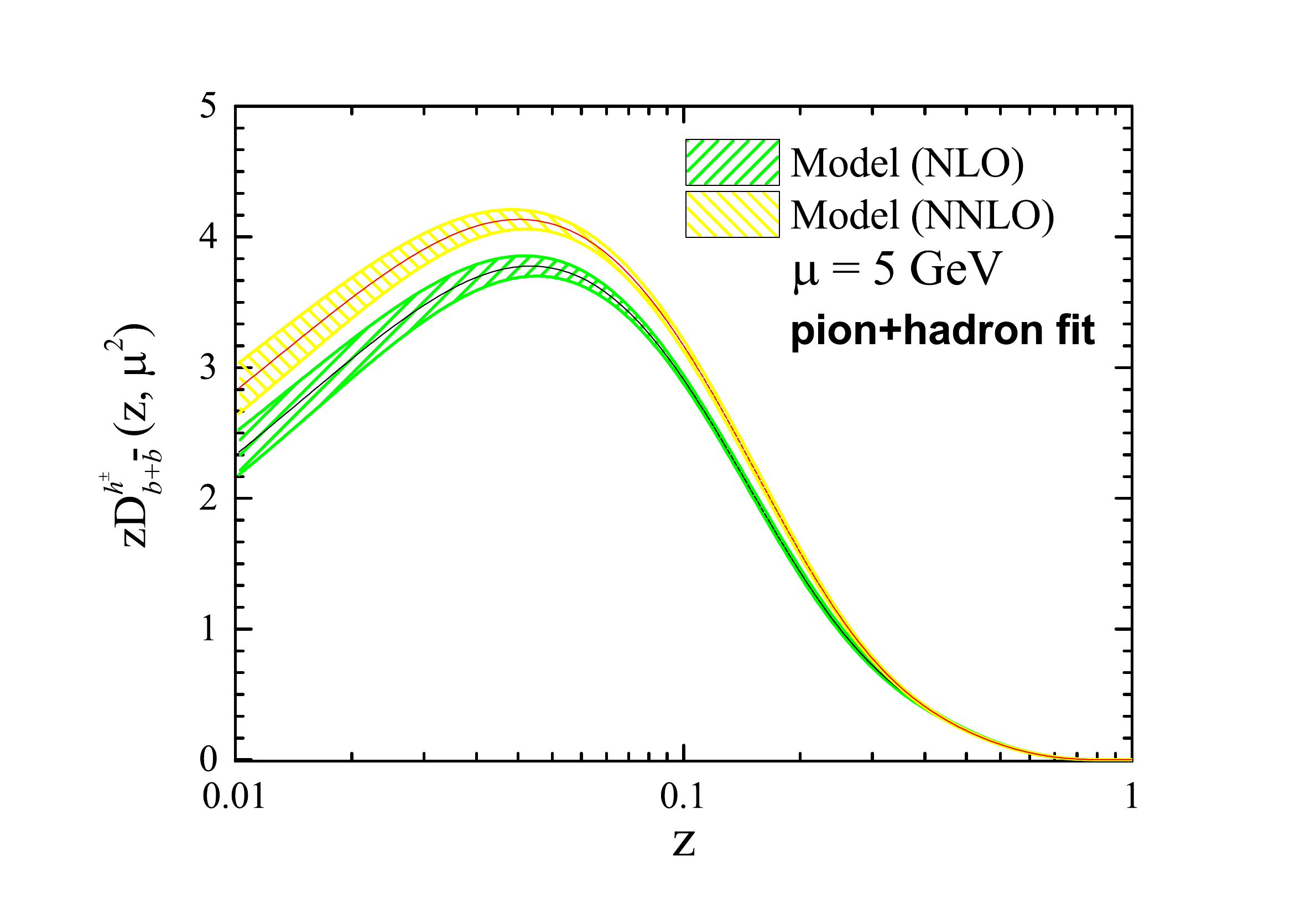}}  	
	\resizebox{0.48\textwidth}{!}{\includegraphics{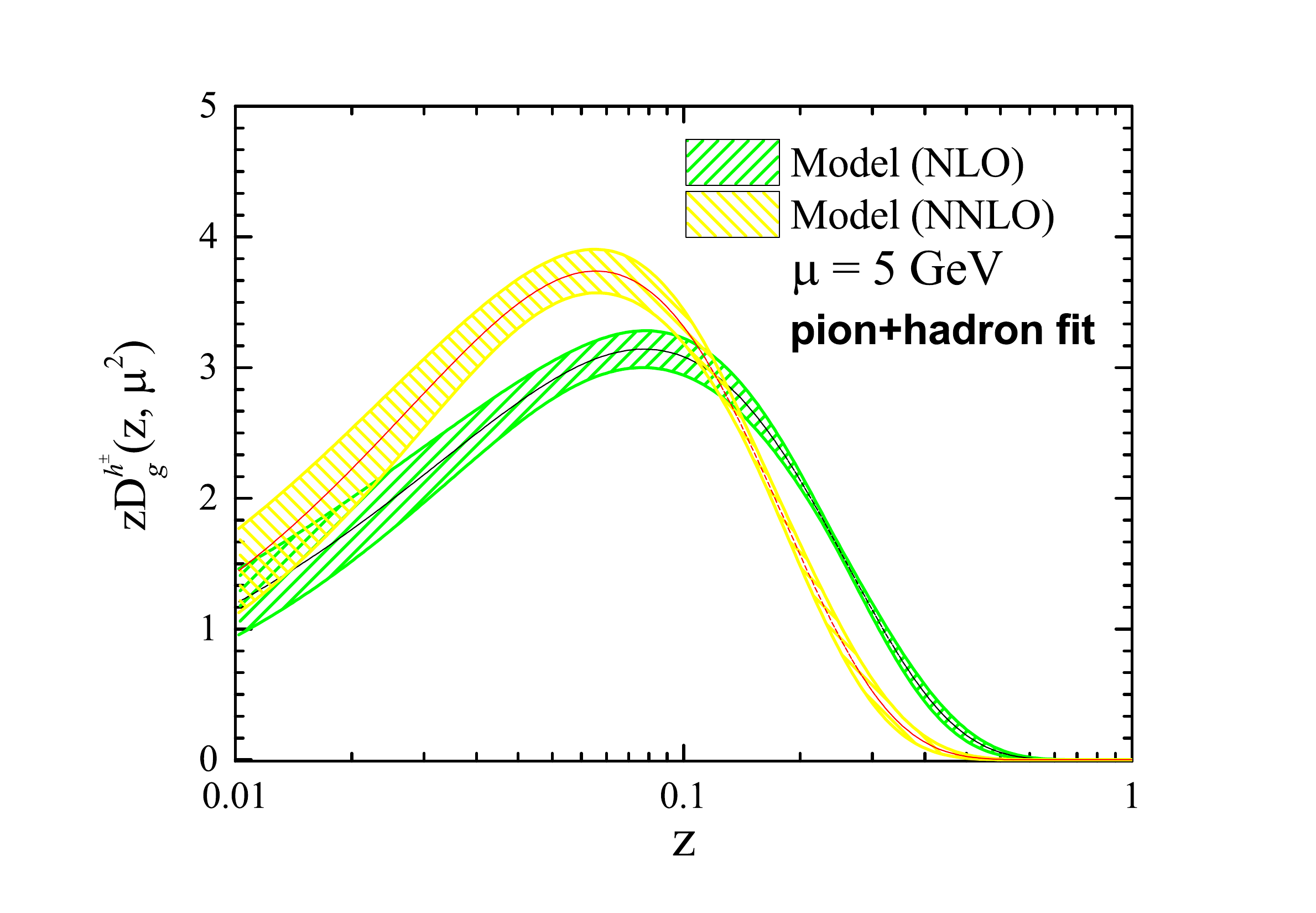}}  	
	\begin{center}
		\caption{{\small  Comparison between the NLO and NNLO pion FFs determined from a simultaneous analysis of pion and unidentified light charged hadron data , ``{\tt pion+hadron fit}", for all flavor distributions at $Q_{0}=5$~GeV. } \label{fig:NLO_NNLO}}
	\end{center}
\end{figure*}

\subsection{Comparison of the ``{\tt pion+hadron fit}" at NLO and NNLO accuracy}

Considering the ``{\tt pion+hadron fit}" analysis as a final and more excellent analysis to determine the pion FFs from SIA data, it is also of interest to compare the distributions obtained at NLO and NNLO accuracy. A comparison between the NLO and NNLO pion FFs determined from a simultaneous analysis of pion and unidentified light charged hadron data for all flavor distributions at $Q_{0}=5$~GeV has been shown in Fig.~\ref{fig:NLO_NNLO}. Overall, we can say that no improvement will be achieved in FF uncertainties by going from NLO to NNLO accuracy. However, there are some crucial changes in the central values of the obtained densities. 
As can be seen, the $u+\bar u$ and gluon FFs follow similar manner. To be more precise, although the size of the changes is not too large, but both of them are increased at smaller values of $z$ and decreased at larger values since the NNLO corrections are included. The $c+\bar c$ and $b+\bar b$ FFs are partially changed just at smaller values of $z$. But the situation is completely different for the case of $s+\bar s$ FF. Actually, the magnitude of its distribution grows to a great extent by considering the NNLO corrections. Note that, although the uncertainty band of $s+\bar s$ FF at NNLO is bigger than NLO one, but the relative uncertainties of two distributions (similar to Fig.~\ref{fig:Ratioq5NLO}) are of the same order.

\subsection{Comparison of the data and theory predictions }

Now we are in a position to complete our study of the fit quality as well as the data vs. theory comparisons.

\begin{figure*}[htb]
	\vspace{0.50cm}
	\resizebox{0.48\textwidth}{!}{\includegraphics{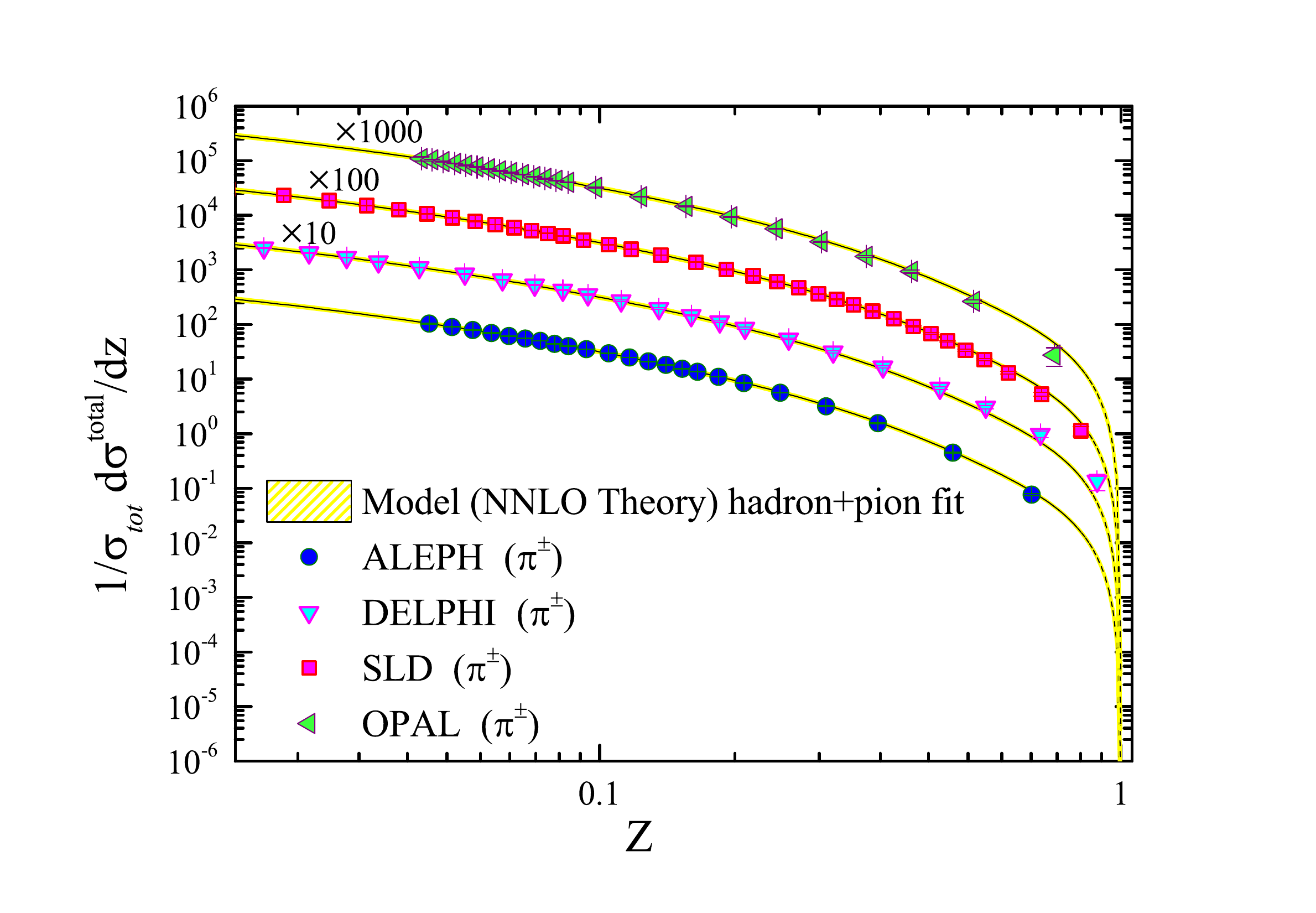}}  	
	\resizebox{0.48\textwidth}{!}{\includegraphics{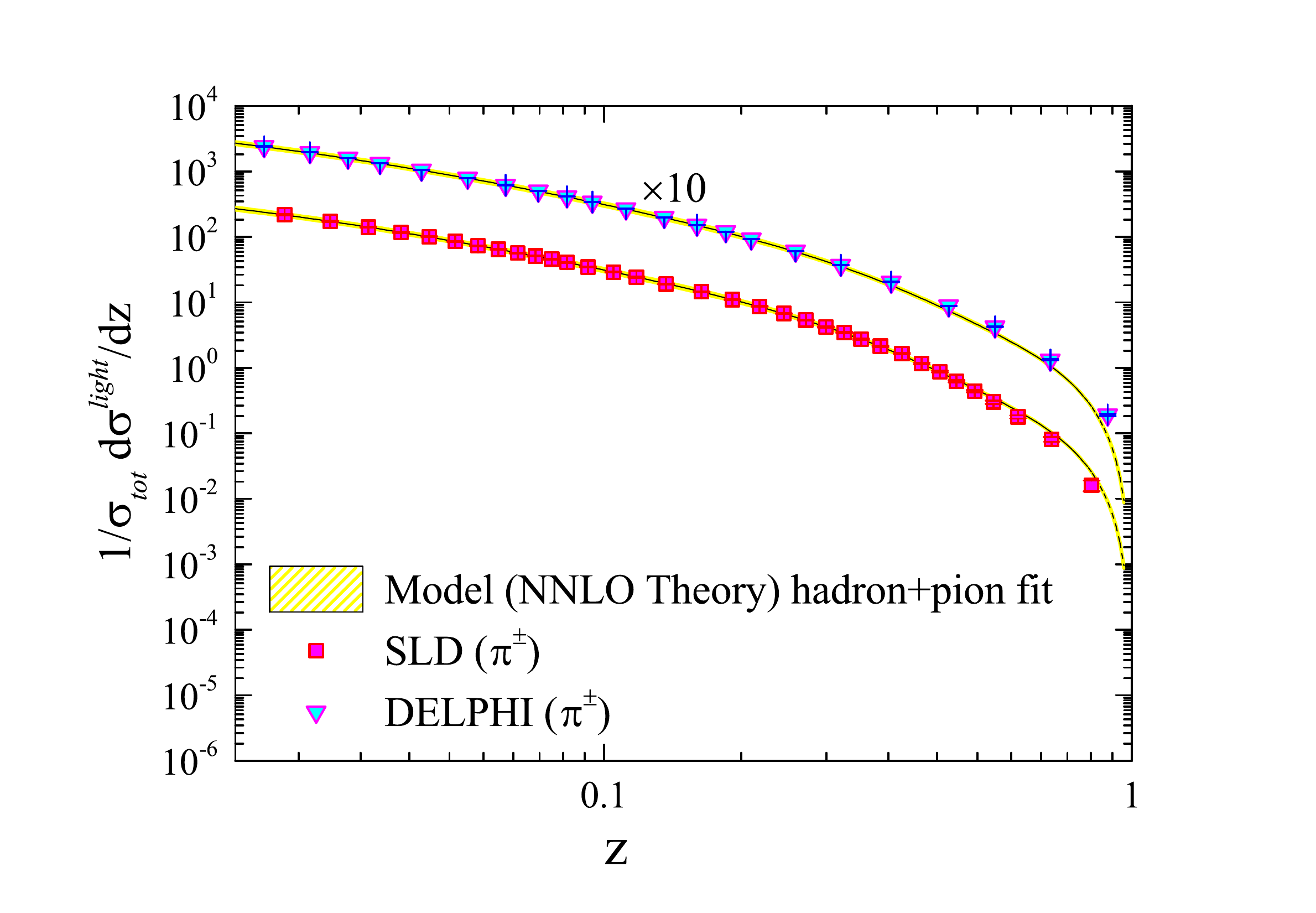}}  	
	\resizebox{0.48\textwidth}{!}{\includegraphics{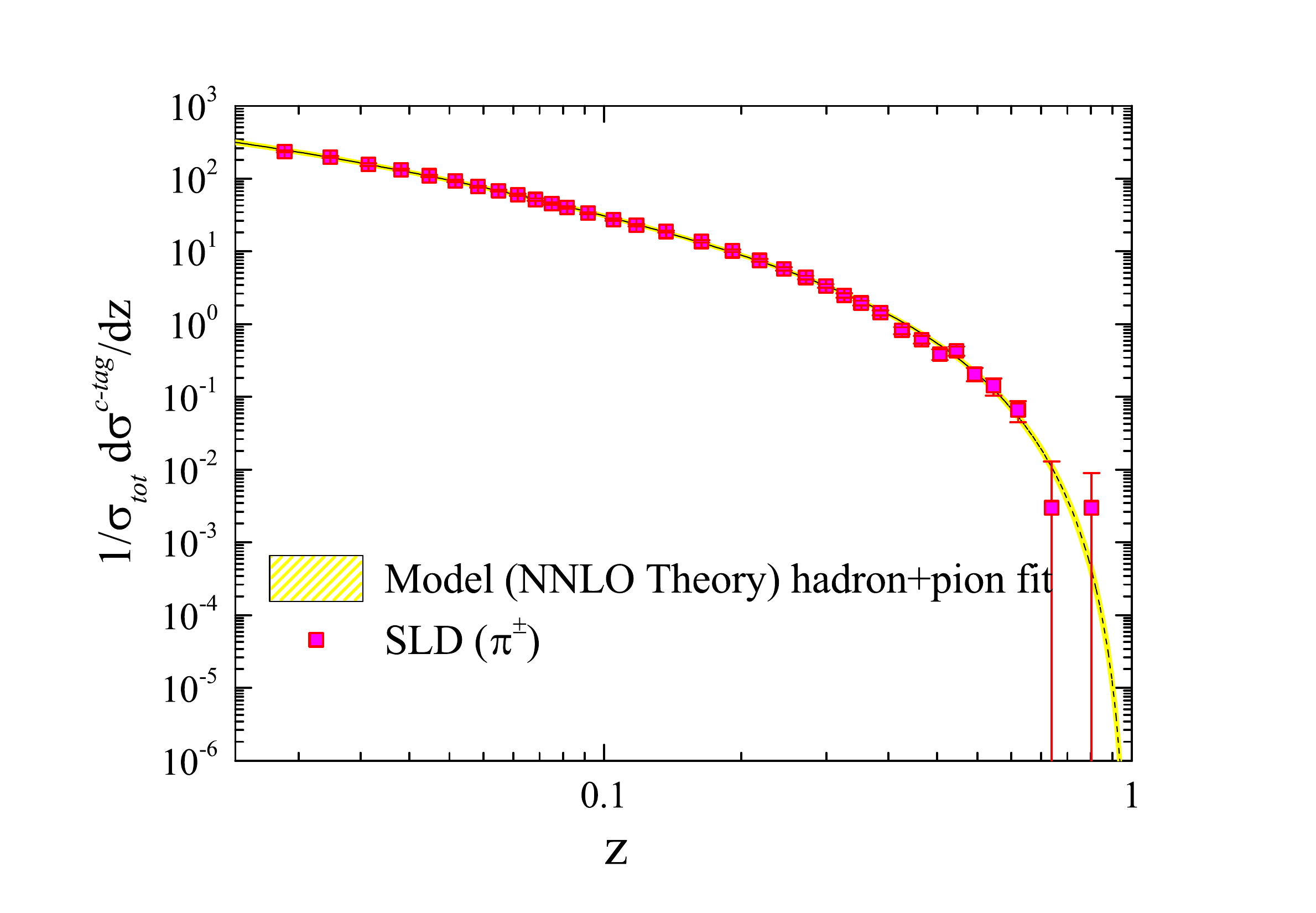}}  	
	\resizebox{0.48\textwidth}{!}{\includegraphics{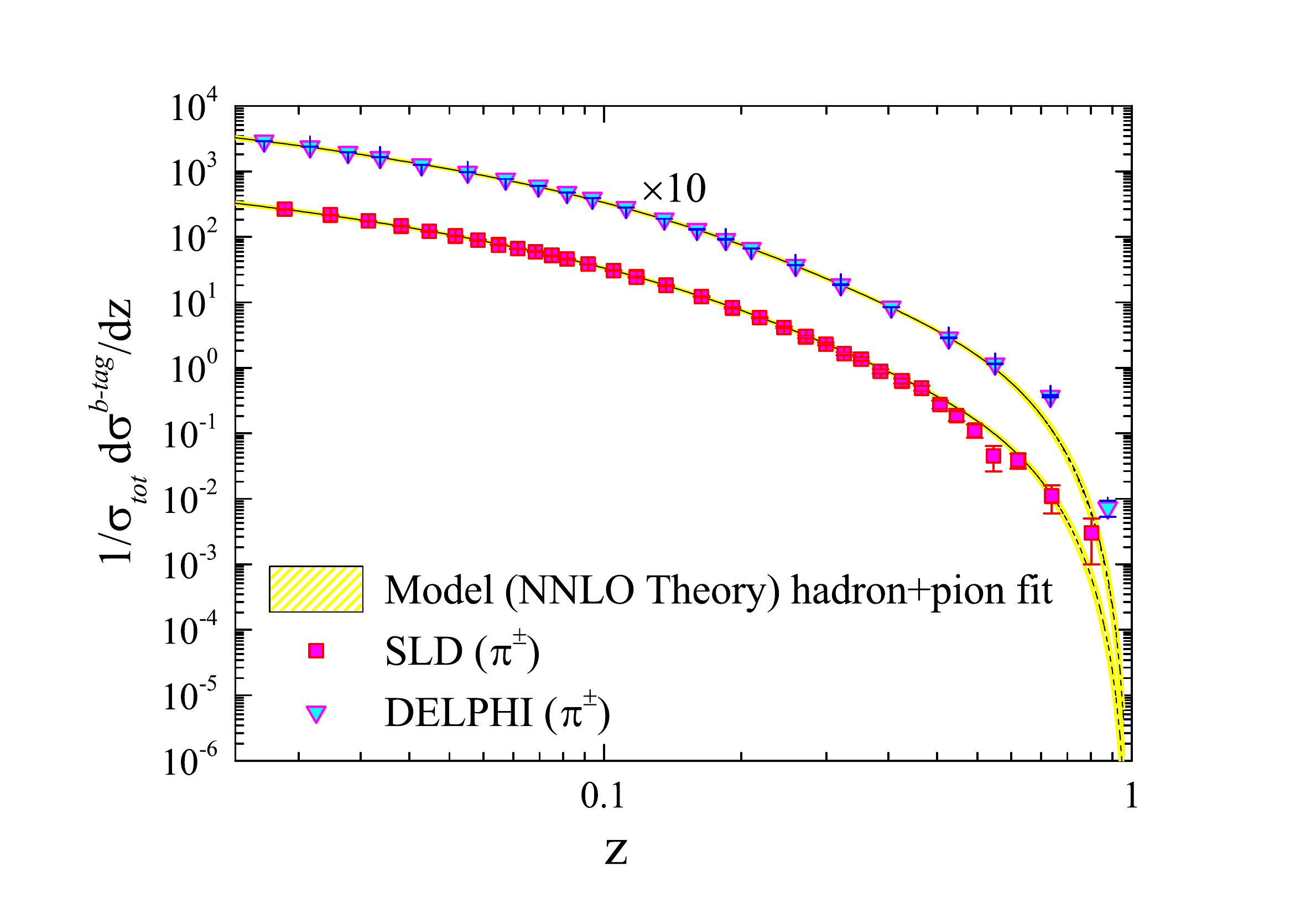}}  	
	\begin{center}
		\caption{{\small Detailed comparisons of $\frac{1}{\sigma_{\it tot}} \frac{d \sigma^{\it \pi^\pm}} {dz}$  with the SIA data sets for the charged pion productions at {\tt ALEPH}, {\tt DELPHI}, {\tt SLD} and {\tt OPAL} experiments. } \label{fig:data_NNLO_Compare}}
	\end{center}
\end{figure*}
\begin{figure*}[htb]
	\vspace{0.50cm}
	\resizebox{0.48\textwidth}{!}{\includegraphics{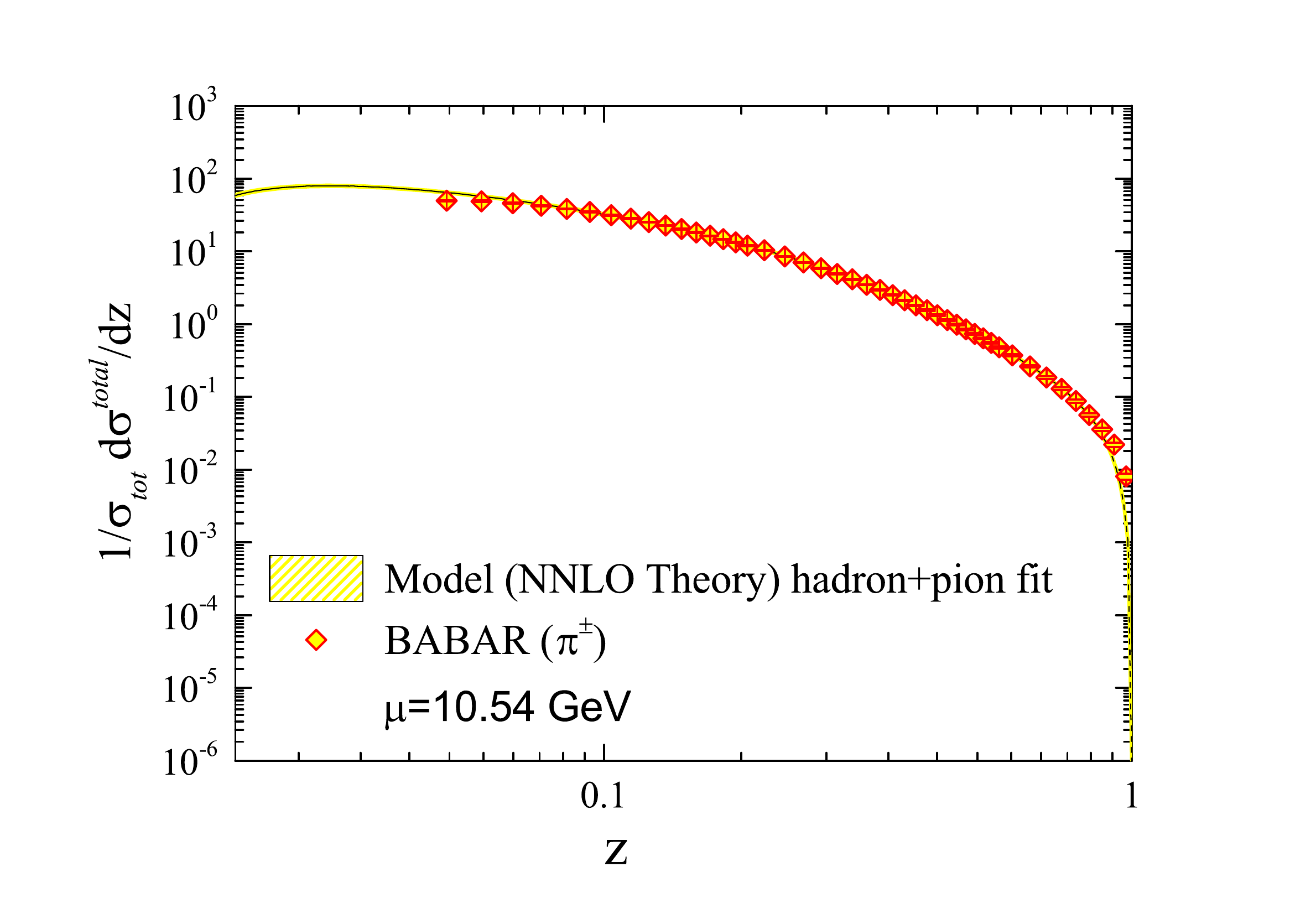}}  	
	\resizebox{0.48\textwidth}{!}{\includegraphics{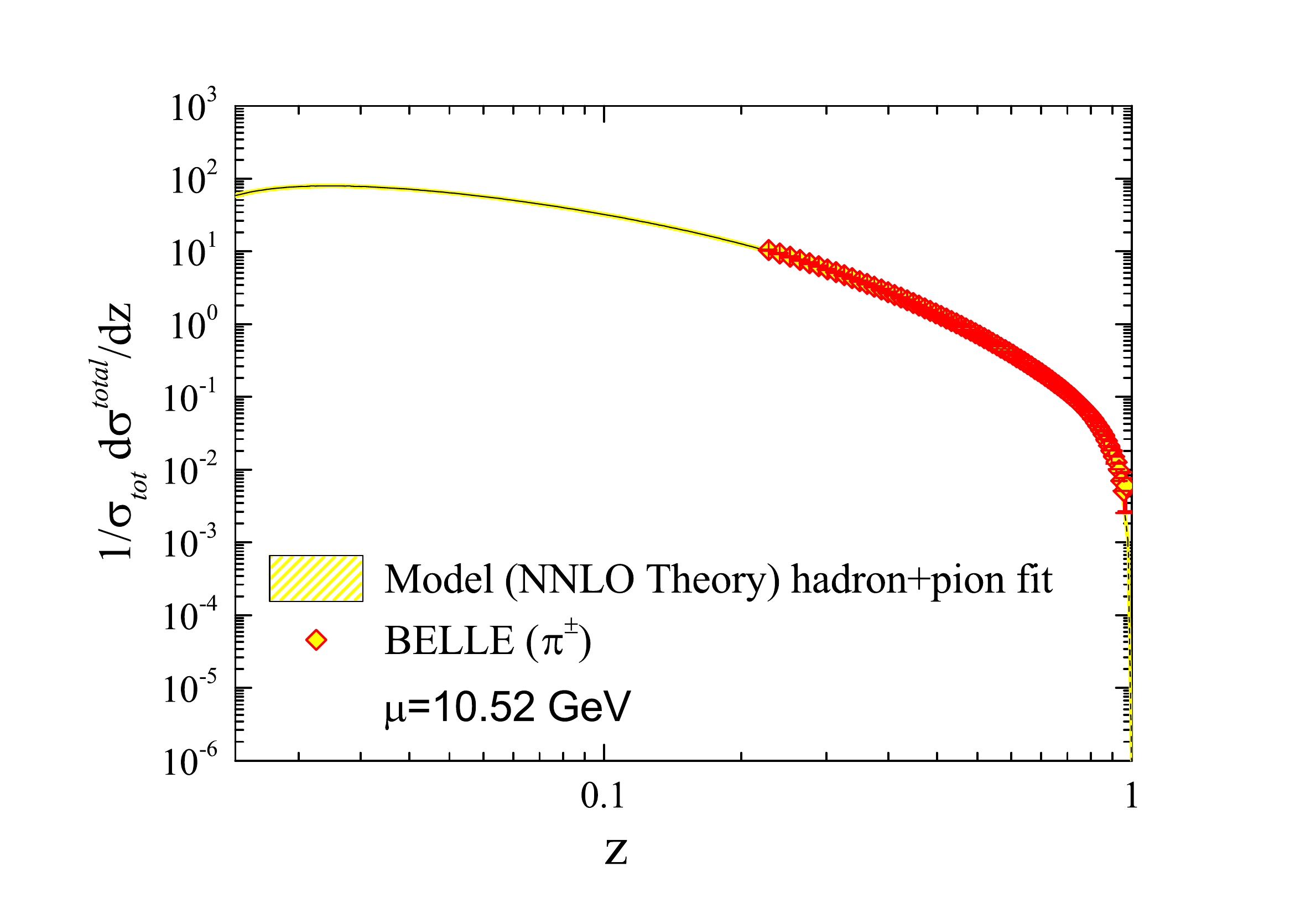}}  	
	\begin{center}
		\caption{{\small Detailed comparisons of $\frac{1}{\sigma_{\it tot}} \frac{d \sigma^{\it \pi^\pm}} {dz}$  with the SIA data sets for the charged pion productions at {\tt BABAR} and {\tt BELLE} experiments. } \label{fig:data_NNLO_BaBar_Belle}}
	\end{center}
\end{figure*}

Here we will focus on the theory prediction based on the extracted pion FFs from our ``{\tt pion+hadron fit}" analysis. We turn to consider only the NNLO results to calculate the normalized cross section for the total, light, $c$-tagged and $b$-tagged. To begin with, in Fig.~\ref{fig:data_NNLO_Compare}, 
we show the detailed comparisons of $\frac{1}{\sigma_{\it tot}} \frac{d \sigma^{\it \pi^\pm}} {dz}$  with the SIA data sets analyzed in this study. These data sets include the charged pion productions at {\tt ALEPH}, {\tt DELPHI}, {\tt SLD} and {\tt OPAL} experiments.  As we can see from this comparison, the agreement between the analyzed data sets and theoretical predictions for wide range of $z$ are excellent, which show both the validity and the quality of the QCD fits. 
In Fig.~\ref{fig:data_NNLO_BaBar_Belle}, we show the comparison between the NNLO theory based on our ``{\tt pion+hadron fit}" with the charged pion productions at {\tt BABAR} and {\tt BELLE} experiments. From the comparisons in this figure, we can see again that the data vs. theory comparisons are excellent. 

As a short summary, considering the impact of these two types of data on the pion FFs, shown in
plots presented in this section, one sees that in the case of ``{\tt pion+hadron fit}" analysis there is a visible reduction on the pion FFs uncertainties at
wide range of $z$, showing that the inclusion of two data sets simultaneously is somewhat more constraining.

%
\section{ Summary and Conclusions } \label{sec:conclusion}
%

In this study, we have quantified the constraints that the unidentified light charged hadron data sets could impose on the determination of pion FFs.
To achieve this goal, new determinations of pion FFs at NLO and NNLO QCD corrections have been carried out based on a comprehensive data sets of SIA processes.
In this respect, we calculate the pion FFs from QCD analyses of two different data sets. Firstly, the pion FFs are determined through QCD analyses of pion experimental data sets alone,  which is referred to as ``{\tt pion fit}".
In addition to the determination of pion FFs using pion experimental data sets, one may certainly expects further constraints to become available for pion FFs studies and an improved knowledge of the FFs will become possible from other source of experimental information. Although the data sets of pion production in electron-positron annihilation include inclusive, $uds$-tagged, $c$-tagged and $b$-tagged observables, some of the parameters of pion FFs at initial scale can not be constrained well enough. Since the most contribution of unidentified light charged hadrons cross sections in SIA measurements is related to the identified pion production, one can expects further constraints by adding these data sets into the QCD fits. 
Hence, to achieve the first and new determination of pion FFs, we have explicitly chosen our input dataset and calculated pion FFs adding simultaneously the pion and unidentified light charged hadron data sets in our analysis, which is entitled as ``{\tt pion+hadron fit}". Our main finding is that using the pion experimental data along with the unidentified light charged hadron data sets has the potential to significantly reduce the pion FFs uncertainties in a wide kinematic range of momentum fraction $z$.

According to the plots presented in this study, one can clearly sees the reduction of pion FFs uncertainties in almost all range of $z$. The most effects of adding unidentified light charged hadron data sets in ``{\tt pion+hadron fit}" analysis are seen for the $s + \bar{s}$ and gluon FFs. Not only the uncertainties of $s+\bar{s}$ and gluon decrease, but also the behavior of their central values have changed considerably. Consequently, our study shows that applying unidentified light charged hadron observables together with pion production data sets in a calculation of pion FFs leads to somewhat a better fit quality.
Since the higher-order corrections are significant, we plan to study the effect arising from higher order correction in the determination of pion FFs. Since we include the SIA data sets in our analyses, the perturbative QCD corrections up to NNLO accuracy can be considered. We found that our results at NNLO corrections improved the fit quality in comparison to the NLO accuracy and it leads to reduction of the $\chi^2$ for all data sets separately as well as for the total $\chi^2$. By considering the NNLO corrections, similar slight improvements in the FFs uncertainty are also found in some region of $z$.

The two analyses presented in this study share, however, a common limitation. In both cases, it has indeed been necessary to include other source of experimental information such as the data from semi-inclusive deep inelastic scattering (SIDIS), and proton-proton and proton-antiproton collisions measured by TEVATRON, RHIC and LHC. However, the NNLO calculations for such processes are not yet available, which would require a relentless effort for the QCD calculations.
It is worth mentioning here that our investigations in this study could be extended to the new determination of kaon and proton FFs considering the unidentified light charged hadron data sets as well as the identified charged hadron production observables. More detailed discussions of these new determination of kaon and proton FFs will be presented in our upcoming study.

%
\begin{acknowledgments}
%

Authors are thankful to Valerio Bertone for many helpful discussions and comments.
Authors thank School of Particles and Accelerators, Institute for Research in Fundamental Sciences (IPM) for financial support of this project.
Hamzeh Khanpour also is thankful the University of Science and Technology of Mazandaran for financial support provided for this research.

\end{acknowledgments}
%



\end{document}